\documentclass{aa}
\usepackage{psfig}
\begin{document}
\thesaurus{08 
i(08.01.1; %Stars: abundances
08.01.3; %Stars: atmospheres
08.05.1; %Stars: early-type
10.01.1; %Galaxy: abundances
10.03.1; %Galaxy: center
10.05.1;) %Galaxy: evolution
}
\title{Chemical abundances in the inner 5\,kpc of the Galactic disk}
\subtitle{}
\author{S.J.~Smartt\inst{1,4}, K.A.~Venn\inst{2}, P.L. ~Dufton\inst{3}\thanks{On leave of absence at the Isaac Newton Group
of Telescopes, Apartado de Correos 368, E-38700, Santa Cruz de La Palma}, 
D.J.~Lennon\inst{4}, W.R.J.~Rolleston\inst{3} and F.P. ~Keenan\inst{3},}
\offprints{S.J.~Smartt}
\institute{  
 {1.}  Institute of Astronomy, University of Cambridge,  Madingley Road, 
 Cambridge, CB3 0HA \\
 {2.} Department of Physics and Astronomy, Macalester College, St. Paul, MN
 55101, USA\\
 {3.}  The Department of Pure and Applied Physics, The Queen's University
 of Belfast, Belfast BT7 1NN, N.~Ireland\\   
 {4.} ING Telescopes, Apartado de Correos 368, Santa Cruz De 
 La Palma, 38780, Canary Islands, Spain.\\}
\date{Received date: ; accepted date: }
\authorrunning{Smartt et al.}
\titlerunning{Abundances in the inner Galactic disk}
\maketitle
\markboth{Smartt et al.}{Inner Galaxy Abundances}

%%%%% AUTHORS - PLACE YOUR OWN MACROS HERE %%%%%
\newcommand{\kms} {km s$^{-1}$\ }
\newcommand{\dk}  {\,dex\,kpc$^{-1}$}
\def\lesssim{\mathrel{\hbox{\rlap{\hbox{\lower4pt\hbox{$\sim$}}}\hbox{$<$}}}}
\def\gtrsim{\mathrel{\hbox{\rlap{\hbox{\lower4pt\hbox{$\sim$}}}\hbox{$>$}}}}

\begin{abstract}

High-resolution, high signal-to-noise spectral data are presented 
for four young B-type stars lying towards the Galactic Centre. 
Determination of their atmospheric parameters from their
absorption line profiles, and $uvby$ photometric
measurement of the continua
indicate that they are massive objects lying slightly out of the
plane, and were probably born in the disk between 2.5$-$5\,kpc from
the Centre. 
We have carried out a detailed absolute and differential line-by-line 
abundance analyses 
of the four stars compared to two stars with very similar atmospheric
parameters in the solar neighbourhood. The stars appear to be 
rich in all the well sampled chemical elements (C, N, Si, Mg, S, Al), 
{\em except for oxygen}. Oxygen abundances derived in the atmospheres of
these four stars are very similar to that in the  solar neighbourhood. 
If the photospheric composition of these young stars is reflective of
the gaseous ISM in the inner Galaxy, then the values derived for the
enhanced metals are in excellent agreement with the extrapolation of
the Galactic abundance gradients previously derived by Rolleston
et al. (2000) and others. However, the data for oxygen suggests that the
inner Galaxy may not be richer than normal in this element, and the 
physical reasons for such a scenario are unclear. 

\end{abstract}
\section{Introduction} 
The current metallicity of the Galactic disk represents a 
fossil record of chemical enrichment 
since the birth of the Milky Way and during its subsequent evolution. 
As stars continually enrich the interstellar medium with their 
nucleosynthetic products, the metallicity of the disk increases
over time. By investigating the relative abundance of chemical elements 
as a function of time and position within the Galaxy, it is possible 
to constrain models of Galaxy formation and evolution. 

Much work has been done on tracking the chemical evolution of the 
solar neighbourhood as a function of time using age-metallicity 
relations and element abundance ratios 
in cool F and G-type stars. The seminal work by
Edvardsson et al. (\cite{Edv93}) showed the potential of using samples of 
nearby
stars with an age spread to trace the temporal evolution of the disk
and the halo, and such work has been extended by, for example, Fuhrmann 
(\cite{fuhr98}) and Israelian et al. (\cite{isr98}). 
These types of metallicity ratios 
have been modelled by many authors (e.g. Carraro et al. 1998, 
Pilyugin \& Edmunds 1996 and references therein), in order to 
constrain our Galaxy's history. The photospheric abundances of 
massive stars (viz. stars with ZAMS $\ga$ 10 M$_{\odot}$) do not 
have the potential to provide us with temporal information as
they are by nature short-lived objects (with typical lifetimes
of the order of $\sim$10-50\,Myrs). However, their high intrinsic 
luminosities mean that we can spectroscopically observe them 
at large distances from the solar neighbourhood; OB-type main-sequence 
stars are typically 
8$^{m}$ visually brighter than early G-type dwarfs. Their 
photospheres are representative of the interstellar material from which they
were born due to their relative youth. 
Hence, we can spatially and radially sample the metallicity of the 
Galactic disk by determining abundances in distant, luminous blue 
stars. The radial variation of different elements provides a further key 
constraint on the formation and evolutionary mechanisms of the 
Galactic disk (Matteucci \& Francois \cite{Mat89}, 
Matteucci \& Chiappini \cite{Mat99}, Prantzos \&\ Aubert \cite{Pra95}, 
Portinari \& Chiosi \cite{por99}, Pagel \& Tautvai\v{s}in\.{e} \cite{pag95}). 

These models require further observational constraints to restrict the 
possible influence of their many free parameters such as initial mass function, 
gas fraction variations, star formation rates, infall and outflow 
rates and metallicity dependent stellar yields. Previous work has shown 
that the metallicity of the Galactic disk decreases as Galactocentric
distance increases. This metallicity gradient has been 
reproduced quantitatively using a number of different methods. 
For example, extensive studies of optical 
emission lines in H\,{\sc ii} regions  by Shaver et al. 
(\cite{Sha83}) and Fich \& Silkey (\cite{Fic91}) have 
derived an oxygen abundance gradient of $-0.07\pm0.02$\,dex\,kpc$^{-1}$; 
within the region $6 \la R_{\rm g} \la 15$\,kpc (where $R_{\rm g}$ refers
to Galactocentric distance throughout this paper). Good agreement 
was also found by Maciel \& K\"oppen (\cite{Mac94}) and 
Maciel \& Quireza (\cite{Mac99}) from 
PN studies, and by Afflerbach et al. (\cite{Aff97}) in a study of 
ultra-compact H\,{\sc ii} regions. Several studies of B-type 
main-sequence stars throughout the disk had given discrepant results, and 
a much shallower gradient than that derived from the nebular studies
(e.g Fitzsimmons et al. \cite{Fit92}, and references therein).
However Smartt \& Rolleston 
(\cite{Sma97}), in a consistent re-analysis of an extensive data-set of B-type 
stars covering $6 \la R_{\rm g} \la 18$\,kpc, showed that previous 
B-type star studies produced spurious gradient determinations, 
and derived an oxygen 
abundance gradient of $-0.07\pm0.02$\,dex\,kpc$^{-1}$ in excellent
agreement with the nebular studies. As a follow up to this, Rolleston 
et al. (\cite{roll2000}; hereafter RSDR) 
carried out a more detailed and exhaustive
survey of the large data-set gathered by the group at Queen's University
and have confirmed this result, while also producing abundance
ratios of different elements along the disk. Further work by 
Gummersbach et al. (\cite{Gum98}) also indicates that the nebular and 
stellar results are in agreement within the errors of measurement. 

However, most studies of the chemical composition of 
the Galactic disk have concentrated on the solar neighbourhood,
and the anti-centre direction. Optical extinction rises steeply
in the direction toward the Galactic Centre, and restricts the 
optical observation of disk stars more than a few kiloparsecs away. 
In the anti-centre direction we have previously probed out to 
$R_{\rm g}\sim$18\,kpc, 
experiencing A$_v\simeq$3 (Smartt et al. \cite{Sma96}), but have 
only managed to observe open clusters towards the Centre approximately 
2$-$3\,kpc away (RSDR) with a reddening of 
A$_v\simeq$2. The metallicity within the solar Galactic radius is poorly
sampled, with only a few IR studies available. 
Rudolph et al. (\cite{Rud97}) and Afflerbach et al. (\cite{Aff97}) have used similar observational methods for
far-IR emission lines in H\,{\sc ii} regions (of differing sizes)
as ISM abundance probes. This far-IR technique has allowed observation
of some inner Galactic regions. These studies suggest that the 
abundance gradients for both nitrogen and sulphur do tend to increase
towards the Galactic Centre, but both indicate that their 
oxygen abundance derivations may suffer some uncertainties (see 
Section\,\ref{discussion_i}). Carr et al.\ (\cite{carr2000}) have 
achieved a first determination of stellar abundances (in an M-type
supergiant) directly at the Galactic Centre in the spectacular
star forming region within the central few hundred parsecs. 
They find an Fe abundance close to that of the solar neighbourhood,
which at first may seems surprising given the wealth of evidence 
of a strong abundance gradient in the Milky Way (and also in other
spirals).   However as we discuss in Section\,\ref{dist_discuss},
this region may be unrelated to the starforming Galactic disk
($3 \la R_{\rm g} \la 18$\,kpc) as a whole, and its evolution 
may have proceeded independently.  

The metallicity of the inner disk therefore has not been investigated 
in any great detail, and lacks information on a variety of elemental
abundances. Spectra of B-type stars provide the potential 
to sample many more chemical elements than is possible through 
emission line studies. As discussed above, massive B-type main-sequence
stars are excellent probes of the current interstellar medium
and sample particularly well the $\alpha$-processed elements (O, Mg, and
Si) as well as C and N (Smartt et al. 1996).  
This paper describes observations of metal-rich 
massive stars lying towards the Galactic Centre together with
detailed model atmosphere and abundance analyses. This allows the 
comparison of abundance ratios in this interesting regime with those
in the rest of the disk, as a key probe of the evolution of our Galaxy.

\section{Observational Data}
The main difficulty in optically probing the inner regions of the Galactic disk is 
of course the high extinction encountered. This section describes how we 
identified targets for optical photometry and spectroscopy which are in 
regions of low extinction, but which have probably originated in the inner disk 
region.

\subsection{Selection of targets in regions of low extinctions towards
the Galactic centre}
\label{target_sel}
The catalogue of Reed (\cite{reed93}) provides a database of 
published UBV$\beta$ photoelectric photometry for objects in the Stephenson 
and Sanduleak catalogue {\em Luminous Stars in the Southern Milky Way} (\cite{step71}). 
This catalogue is a fairly complete record of OB-type stars and 
A- \& F-type supergiants and giants within 10$^{\circ}$ of the galactic 
plane and with longitude limits $-174^{\circ} < l < +81^{\circ}$, down 
to a limiting photographic magnitude of approximately 12.5.
In order to identify OB-type stars near the Galactic Centre, which also have 
moderate Galactic latitudes and hence suffer from low extinction,  
stars from the Reed catalogue were selected which satisfied all of the 
criteria listed 
below:--
\begin{enumerate}
\item Stars within $-40^{\circ} < l < +20^{\circ}$ ; hence restricting the 
sample to objects towards the Galactic Centre. 
\item Stars which are more than 4$^{\circ}$ above or below the plane; in order
to limit the effects of interstellar extinction. 
\item Stars which are fainter than V = 11.0, to sample the more distant
objects. 
\item Stars with a colour factor between $-$0.90 $<$ Q $<$ $-$0.66; 
appropriate for early B-type stars;\\ 
\mbox{where  $Q = (U - B) - 0.72(B - V)$}. 
\item Stars with a B--V of less than 0.5 $-$ which implies a reddening of
$E(B-V) \la 0.8$ (for early B-type intrinsic colours). 

\end{enumerate}

The combination of these criteria should produce targets which are 
potentially normal early B-type dwarfs (or {\em near} main-sequence objects)
lying within $\sim$4\,kpc of the Galactic Centre and twelve such stars 
were found. 

The database of Reed \& Beatty (\cite{reed95}) contains 
published spectral types for a subset of the objects in 
Reed (\cite{reed93}). However, the fainter objects of 
the earlier study (which were the most promising distant B-type stellar
candidates) did not have an entry in Reed \& Beatty (1995). 
As a result we had to rely on colours as a primary selector. 

\subsection{Spectroscopic data}
The spectroscopic data presented in this paper were taken 
on two different runs. All twelve candidates
were observed at the Anglo-Australian Telescope 
in August 1994
at intermediate resolution. The RGO spectrograph along with the 1200B 
diffraction grating, 82\,cm camera and the TEK\,1K CCD were used to give a
resolution of approximately 0.7\,\AA~ (FWHM). Two grating positions were 
employed with central 
wavelengths of 4020\,\AA~ and 4580\,\AA,  each giving a wavelength coverage of 
240\,\AA. The primary motivation of these observations was to identify normal
stars for further, higher resolution observations at a higher signal-to-noise
ratio, and to provide reliable H\,{\sc i} profiles for the stars which did turn
out to be normal.  Six of the stars observed appeared to be normal 
main-sequence early-type B-stars. However one of these was a fast rotator
with a lack of metal line features, while another appeared to be a 
double-lined spectroscopic binary (composed of two early-type components)
with the metal lines severely blended. The nature of the other six peculiar 
stars is discussed in Venn et al. (\cite{ven98}). 
The four apparently normal stars (see Table~\ref{photometry_tab1})
were further observed with the 
ESO 3.6m telescope using the CASPEC \'{e}chelle spectrograph during July 
1995. The central wavelength was 4500\,\AA~ with
a wavelength coverage of near 1300\,\AA~ (3920 -- 5240\AA)
at a resolution of $\sim$0.1\,\AA. 

\begin{table}
\begin{center}
\caption[]{Photometric details of Galactic Centre B-type main-sequence stars.
Our estimated spectral types are quoted}
\label{photometry_tab1}
\vspace{0.25cm}
\begin{tabular}{lcccccc} \hline
Star     & SpT         &   V   & B-V  & U-B   & $l^{\circ}$ & $b^{\circ}$\\
\hline
LS5130   & B2\,IV      & 12.09 & 0.24 & $-$0.53 & 21.12 & $-$5.63  \\
LS4419   & B2\,IV      & 11.08 & 0.08 & $-$0.65 & 351.78  & $-$5.88       \\
LS4784   & B2\,IV      & 11.55 & 0.09 & $-$0.60 & 1.70  & $-$6.12       \\
LS5127   & B3\,II-III  & 11.96 & 0.15 & $-$0.55 & 16.13 & $-$7.34       \\
\hline
\end{tabular}
\end{center}
\end{table}

During both
observing programmes the observational routine was similar. Each star
was observed long enough to accumulate sufficient counts in the continuum
to yield a minimum signal-to-noise ratio of $\sim$70, and in most cases  
more than 100. The observations were split into exposures of 
1200 -- 1500\,s to minimize the impact of cosmic-ray contamination and 
were bracketed by observations of Cu-Ar and/or Cu-Ne wavelength calibration
lamps. Flat-field exposures and bias frames were generally taken at both the
beginning and the end of the night. 

The single order spectra were reduced using the {\sc starlink} 
package {\sc figaro} (Shortridge et al. \cite{short97}) as discussed in 
Smartt et al. (\cite{Sma96}), where further details can be found. 
The \'{e}chelle data were reduced using the Image Reduction and 
Analysis Facility ({\sc iraf}), using standard techniques. When in one
dimensional format, all the spectra were transferred for further
analysis to the {\sc starlink} program {\sc dipso} (Howarth et
al. \cite{How98}). Normalization was achieved by carefully selecting
continuum regions free from absorption lines, and fitting low order (of
degree 3 or 4) polynomials through the noise.  Equivalent widths for
the metal lines and non-diffuse lines of neutral helium were measured by
the non-linear least square fitting of single or multiple Gaussian
profiles to the normalised spectra. Each line is assigned an error
estimate reflecting the reliability of the equivalent width
measurement, viz. a: error less than 10\%, b: error less than 20\%, c:
error greater than 20\%. These were assigned by considering the
numerical error returned by the {\sc dipso} line fitting computation
and the qualitative accuracy of the profile fit.  The hydrogen and
diffuse helium lines were not measured in this manner, but the
normalised profiles were extracted directly for comparison with
Galactic standards and with theoretical profiles. The equivalent width
measurements for each star can be found in Appendix A (available
electronically).

\subsection{Photometric Data}
Str\"{o}mgren $uvby$ photometry of the four B-type LS stars was obtained
in service mode using the People's photometer on the SAAO 0.5-m telescope.
Details of the instrumental set-up can be found in Kilkenny et al. 
(\cite{kil88}) -- and the resultant photoelectric aperture photometry is 
listed in Table\,~\ref{photometry_tab2}. Additionally, we attempted 
to obtain Str\"{o}mgren CCD $uvby$ data (as a consistency check) 
on the ESO 1.5-m Danish Telescope in July 1998. For two of the stars 
(viz. LS4419 and LS5127) we did indeed obtain good quality observations 
during photometric conditions. However, the data for the other two stars 
was compromised by high cirrus. Initial CCD reductions were performed 
using the {\sc iraf ccdred} package (Massey \cite{mas92}), while subsequent
digital aperture and point-spread function photometry was undertaken using 
tasks within the {\sc iraf} package {\sc daophot} (Davis \cite{dav94}). 
Further details of these methods can be found in Mooney et al. (\cite{moo00}). 

In Table\,~\ref{photometry_tab2} we list the individual photometric indices 
as derived from the SAAO observations. Typical uncertainies in the 
SAAO photoelectric photometry are 0.01 and 0.005 magnitudes in colours 
and the individual passbands respectively. The Str\"omgren reddening free 
[$u-b$] index should be accurate to $\pm$0.03 magnitudes. The ESO CCD 
dataset comprised four independent observations per filter per field 
containing each of the LS stars. Photometric errors were estimated using 
two methods. First, each set of $uvby$ instrumental magnitudes were 
independently transformed to the standard system of $y$ and ($b-y$), 
$m_1$, $c_1$ indicies. These were used to deduce mean values and associated 
standard errors of the mean. Secondly, values of the arithmetic mean and 
standard error were deduced for the independent sets of $uvby$ instrumental 
measurements. The mean values were transformed to the standard magnitude 
and colour indicies, while the associated photometric uncertainties were 
deduced by propagating the standard error of the mean values through the 
transformation equations.

For LS4419, both these methods inferred a photometric accuracy of 0.006 
magnitudes in $y$ and 0.011 magnitudes in the colour ($b-y$). The 
reddening free [$u-b$] index should be accurate to $\pm0.02$ magnitudes. 
Thus, it is not surprising that we find excellent agreement between the 
SAAO and ESO photometry for LS4419. Photometric uncertainties are 
somewhat larger for the ESO data obtained for LS5127, viz. 0.015, 0.035 
and 0.15 magnitudes in $y$, colour and [$u-b$] index respectively. For this 
star, the SAAO data should be the more reliable. However, it should be noted 
that the SAAO and ESO measurements agree within the photometric errors -- 
which in turn lead to similar estimates of the stellar effective temperature 
(see Sect.~3).

\begin{table*}
\caption[]{Str\"{o}mgren photometric data for stars observed at the SAAO 0.5m, and the 
ESO 1.5m Danish. Values for $\gamma$~Peg and $\iota$~Her were taken from the
{\sc SIMBAD} database and references therein.}
\label{photometry_tab2}
\vspace{0.25cm}
\begin{tabular}{lcrrcccc}\hline
Star   & $y$  & $(b-y)$ & $m_{1}$ & $c_{1}$ & [$c_{1}$]             & [$u-b$] (SAAO)  & [$u-b$] (ESO) \\\hline
\\
 LS5130  & 12.038 &  0.264  & $-$0.025 &  0.182 & 0.129   & 0.248 & -- \\
 LS4419  & 11.089 &  0.115  & 0.032  & 0.105   & 0.082    & 0.220 & 0.204  \\
 LS4784  & 11.519 &  0.157  & 0.035  & 0.122   & 0.091    & 0.261 & -- \\
 LS5127  & 11.883 &  0.193  & 0.023  & 0.233   & 0.194    & 0.364 & 0.456 \\ 
\\
$\gamma$\,Peg &   2.84    & $-$0.106  & 0.093 & 0.116  & 0.137 &  \multicolumn{2}{c}{0.255}        \\
$\iota$\,Her  & 3.80  &  $-$0.064 & 0.078 & 0.294  & 0.307 &   \multicolumn{2}{c}{0.423}     \\\hline

\end{tabular}
\end{table*}

\section{Model atmosphere and abundance analyses}

The methods employed to derive stellar
atmospheric parameters are similar to those described in Smartt et
al. (1996) and Rolleston et al.  (1997). All results are based on the
ATLAS9 grid of line-blanketed model atmospheres of Kurucz
(\cite{Kur91}). LTE line formation  codes were used to 
derive line profiles leading to determinations of atmospheric parameters 
and chemical compositions. 

\begin{table*}
\caption[]{Atmospheric parameters of the Galactic Centre main-sequence
B-type stars and the bright spectroscopic standards used in the differential
analysis. 
$T_{\rm eff}$ was derived from Str\"{o}mgren photometry; 
typical random errors on the 
derived values are $\pm$1000\,K and $\pm$0.2\,dex for $T_{\rm eff}$ and 
$\log g$, respectively, and $\pm$3\,kms$^{-1}$ for $\xi$. 
For LS4419 and LS5127, the values for $T_{\rm eff}$ are
the mean of the results from the SAAO and ESO data -- and the 
others are derived from the SAAO photometry only. }
\label{Bs_atmos_params}
\vspace{0.25cm}
\begin{tabular}{lccccc}\hline
Star    &  $T_{\rm eff}$ & $\log g$ & $\xi$        & [He]  & v$sini$ \\
        &   (K)               & (cgs)    & (kms$^{-1}$) & dex & (kms$^{-1}$) \\\hline
LS5130  & 21200 & 3.5      &  5           & 11.05 $\pm$0.20 & 35 \\
LS4419  & 22100 & 3.7      &  7           & 10.92 $\pm$0.10 & 80\\
LS4784  & 20900 & 3.9      &  5           & 10.92 $\pm$0.15 & 93\\
LS5127  & 17900 & 3.2      &  7           & 10.87 $\pm$0.13 & 120\\
$\gamma$\,Peg & 21000 &  3.8     &  5           & 10.87 $\pm$0.09 & 5\\
$\iota$\,Her  & 17700 &  3.9     &  5           & 10.78 $\pm$0.08 & 10\\
\hline
\end{tabular}
\end{table*}

\begin{table*}
\caption[]{Absolute elemental abundances for the star LS5130, together
with a line-by-line differential analysis with the ``standard'' star
$\gamma$~Peg which is a chemically normal star lying within the 
solar neighbourhood. The abundances quoted are the mean values from
results from $n$ lines in each star (where $n$ is the number in brackets
after the abundance), and the uncertainties are the sample standard 
deviations. Also listed are absolute abundances normally
associated with bright B-type stars within approximately 500\,pc of
the Sun$^{1}$; and values from ISM or H\,{\sc ii} region studies
in the solar neighbourhood$^{2}$. For reference, the {\em actual}
solar photospheric abundances are listed, and we note (as others have indicated e.g.
Snow \& Witt \cite{sno96}) that the Sun appears significantly metal-rich
compared to the surrounding ISM and neighbouring Population\,I objects (i.e.
young, massive stars) in many elements.}
\label{LS5130_abundances}
\vspace{0.25cm}
\begin{tabular}{llllllll}\hline
Species          &       LS5130        & $\gamma$\,Peg    & Differential  & & Nearby   & Other solar      
&  Solar \\
                 &                    &                   &               & & B-stars$^{1}$ & neighbourhood 
&    \\
\hline 
C\,{\sc ii}   &  8.47 $\pm$ 0.31 (11)  & 8.22 $\pm$ 0.31 (9)  &  +0.23 $\pm$ 0.15 (9)    & ~~~~ & 8.20     
& 8.13$^{c}$  & 8.55 \\
N\,{\sc ii}   &  7.96 $\pm$ 0.22 (25)  & 7.79 $\pm$ 0.24 (19) &  +0.15 $\pm$ 0.13 (19)   & ~~~~ & 7.81     
& 7.94$^{d}$  & 7.97 \\
O\,{\sc ii}   &  8.69 $\pm$ 0.19 (43)  & 8.72 $\pm$ 0.32 (81) &  $-$0.06 $\pm$ 0.16 (35) & ~~~~ & 8.68     
& 8.70$^{c}$  & 8.87 \\
Mg\,{\sc ii}  &  7.64~~~~~~~~~~\,\,(1)   & 7.22~~~~~~~~~~\,\,(1)  &  +0.42~~~~~~~~~~\,\,(1) & ~~~~ & 
7.38$^{a}$  &  --       &  7.58 \\
Al\,{\sc iii} &  6.30 $\pm$ 0.09 (4)   & 6.14 $\pm$ 0.05 (3)  &  +0.13 $\pm$ 0.10 (3)    & ~~~~ &  6.45    
&  --         &  6.47 \\
Si\,{\sc ii}  &  6.85 $\pm$ 0.14 (4)   & 6.50 $\pm$ 0.10 (4)  &  +0.34 $\pm$ 0.04 (4)    & ~~~~ &  
7.28$^{a}$    &  --         & 7.55 \\
Si\,{\sc iii} &  7.66 $\pm$ 0.19 (5)   & 7.29 $\pm$ 0.20 (5)  &  +0.36 $\pm$ 0.24 (5)    & ~~~~ &  
7.28$^{a}$    &  --         & 7.55 \\
S\,{\sc ii}   &  7.39 $\pm$ 0.21 (2)   & 6.96 $\pm$ 0.24 (15) &  +0.38 $\pm$ 0.02 (2)    & ~~~~ & 7.21     
& 7.12$^{d}$  &  7.21 \\
S\,{\sc iii}  &  7.29~~~~~~~~~~\,\,(1)    & 7.06 $\pm$ 0.41 (9)  &  +0.23~~~~~~~~~~\,\,(1)     & ~~~~ &  7.21 
& 7.12$^{d}$ & 7.21 \\
Ar\,{\sc ii}  &  7.03 $\pm$ 0.61 (3)   & 6.75 $\pm$ 0.46 (3)  &  +0.28 $\pm$ 0.15 (3)    & ~~~~ & $-$
& --       &  6.58 \\
Fe\,{\sc iii} &  6.53 $\pm$ 0.96 (3)   & 6.39 $\pm$ 0.82 (3)  &  +0.13 $\pm$ 0.18 (3)    & ~~~~ & 
7.36$^{a}$     
& --        & 7.51 \\
\hline
\end{tabular}
\begin{scriptsize}
\begin{enumerate}
\item All typical B-type stellar values are taken from Gies \& Lambert (1992), 
apart from \\
~~~~ a: Kilian (\cite{Kil94}) \\
\item Solar neighbourhood values taken from :\\
~~~~ c: Local ISM abundances from Meyer et al. (\cite{mey98}), Cardelli et al. (\cite{card96}).  \\
~~~~ d: Orion nebular abundances from Baldwin et al. (1991) \\
\item Solar values taken from Grevesse \& Noels (\cite{gre93}) and Anders \& Grevesse (\cite{and89}). 
\end{enumerate}
\end{scriptsize}
\end{table*}

As a first attempt to constrain the effective temperatures ($T_{\rm eff}$)
of the four LS stars, which were all of spectral type B2/B3, 
we used the ionization balance of the two stages of silicon; 
the Si\,{\sc ii} lines at 4128\,\AA~ and 4130\,\AA, and the Si\,{\sc iii}
triplet at 4552--4574\AA. 
This method has been previously employed to analyse the 
bright standard $\gamma$\,Pegasi (HR39), and is known to produce
anomalous and unsatisfactory 
results (Ryans et al. 1996, Gies \& Lambert 1992, Peters 1976). Absolute 
abundances (derived using this temperature diagnostic) of N\,{\sc ii}, 
O\,{\sc ii}, Mg\,{\sc ii}, and S\,{\sc iii} show systematic 
underabundances of around 0.4\,dex with respect to normal B-type
stellar values.
Further, and more critically, large discrepancies then exist between the 
abundances derived from other ionization equilibria 
(C\,{\sc ii}/C\,{\sc iii}, S\,{\sc ii}/S\,{\sc iii}, and 
Si\,{\sc iii}/Si\,{\sc iv}) when using this Si\,{\sc ii}/Si\,{\sc iii}
temperature balance. 
Using Str\"{o}mgren photometry as an alternative 
effective temperature indicator produces more satisfactory results; 
typically, a value systematically 
3000\,K less than that estimated from the Si\,{\sc ii}/Si\,{\sc iii}
balance has been obtained which agrees with the other ionization balances, 
and the metal abundances give normal results. 
The problem with the Si\,{\sc ii} lines historically stretches back
more than two decades, when Peters (\cite{peter76}) first reported the 
problem. All stars in the Gies \& Lambert (1992) study (which have 
$T_{\rm eff} \sim 21000$; from Str\"{o}mgren indices) appear to produce spurious
Si\,{\sc ii} results and a similar temperature problem has been found
by the authors in 
at least two other stars -- 22\,Ori and $\iota$\,Her. 
In each case the temperature
found from the Si\,{\sc ii}/Si\,{\sc iii} lines is systematically 
higher than that produced from the Str\"{o}mgren indices.
Hence, it appears that the 
Si\,{\sc ii}/Si\,{\sc iii} ionization balance is flawed as a temperature 
diagnostic due to reasons which are, as yet, not well understood. 
We find that including non-LTE corrections 
(Brown 1987) in the line formation calculations fails to remove the discrepancy.
Cuhna \& Lambert (\cite{Cuh94}) have also reported the same effect 
when adopting Str\"{o}mgren temperatures for Orion B-types and 
their subsequent determination of Si abundances from both ionization 
stages. 

In the rest of this paper we shall adopt the effective temperatures 
derived from the 
Str\"{o}mgren indices listed in Table\,~\ref{photometry_tab2}.
The colour index [$u-b$] measures the Balmer jump as a function of 
$T_{\rm eff}$ and we have primarily used the calibration of
Napiwotzki et al. (\cite{nap93}). We also compared the 
effective temperatures derived using 
the calibration of Lester et al. (\cite{Les86}), 
which provides the [$u-b$] vs $T_{\rm eff}$ relation as 
a function of surface gravity; consistent results were found
in each case. Table\,~\ref{Bs_atmos_params} lists the 
effective temperatures that we have adopted. For the two 
stars for which we have both SAAO and ESO data, we list the 
mean of the two temperatures. Excellent agreement was found from the 
two data sets for LS4419, with temperatures of 21900 and 22300\,K derived
from ESO and SAAO respectively; for LS5127 the temperatures were 
in poorer agreement, 18700 and 17100\,K, but consistent 
with a 17900 $\pm$1000\,K range. 

Logarithmic surface gravities ($\log g$) were estimated by fitting 
model profiles of the Balmer lines from our atmospheric analysis codes
to observed H$\epsilon$,
H$\delta$, and H$\gamma$ depending on which lines were available in the 
RGO spectra (see Section\,2). In all cases the results from each 
line were in good agreement. A conservative estimate of 
the probable random errors in our analyses, given the errors in the 
measured data are $\pm$1000\,K and $\pm0.2$\,dex.
A value for microturbulent velocity ($\xi$) was determined for all the B-type
stars by requiring that the derived abundances from, primarily, 
O\,{\sc ii} were independent of line strength. The O\,{\sc ii} ion 
was chosen as it has the richest absorption line spectra in stars of
this type. However where a significant number of N\,{\sc ii} lines 
were available, $\xi$ was independently estimated and the mean taken. 
Atmospheric parameters for all the stars are listed in Table\,~\ref{Bs_atmos_params}. 

As a final step in defining the model atmosphere appropriate to 
each star, we determined the helium abundance within this iterative 
procedure.  
Helium abundances were calculated from the line strengths of the available
non-diffuse lines (3964, 4437, 4713, 5015 \& 5047\,\AA) and the profile 
fitting of the diffuse lines (4009, 4026, 4387, \& 4471\,\AA). In each 
star we found a normal helium composition of approximately $10.9$\,dex 
(corresponding to $y\simeq0.1$ by fraction). Hence, we did not need to 
repeat the $T_{\rm eff} - \log g - \xi $ iterations for models with 
significantly abnormal helium fractions. 
 
Final model atmospheres with parameters which satisfied the above criteria
were then calculated in order to allow the metal abundances from each
unblended ionic feature to be determined in each star. The elemental 
abundance associated with a particular line was derived by 
comparing the observed line strengths and profiles to those predicted from 
the LTE line formation calculations. 
A line-by-line differential analysis was performed for each
star relative to nearby (i.e. within the solar neighbourhood) 
spectroscopic standard stars of very similar
atmospheric parameters. 
The atomic data used in the line formation theory
were from Jeffery (\cite{Jef91}), but this choice is not 
critical for our purposes of a differential abundance analysis.  
Tables~\ref{LS5130_abundances}-\ref{LS5127_abundances} list the 
absolute and differential abundances derived for each star. 
The mean absolute abundances are on a logarithmic scale with 
hydrogen = 12.0\, dex (i.e. [X]$= 12 + \log N_{\rm X}/N{\rm _H}$), 
and the errors quoted are the standard deviations about this mean value.

\begin{table*}
\caption[]{Absolute elemental abundances for the stars LS4419 and LS4784, 
together with a 
line-by-line differential analysis with the ``standard'' star $\gamma$~Peg
which is chemically ``normal'' with very similar atmospheric
parameters.  The absolute abundances for $\gamma$~Peg are shown in the 
previous table.}
\label{LS4419_abundances}
\vspace{0.25cm}
\begin{tabular}{llllll}\hline
Species          & LS4419 {\em (absolute)}  & LS4419 {\em (differential)}  &  & LS4784 {\em (absolute)}  & LS4784 {\em (differential)} \\
\hline 
C\,{\sc ii}      &  8.18 $\pm$ 0.27 (5)    &  +0.07 $\pm$ 0.19 (5)     & ~~~   & 8.42 $\pm$ 0.56 (4)  & +0.27 $\pm$ 0.16 (4)   \\
N\,{\sc ii}      &  8.27 $\pm$ 0.52 (16)   &  +0.43 $\pm$ 0.25 (13)    & ~~~   & 8.43 $\pm$ 0.37 (15) & +0.62 $\pm$ 0.29 (12)   \\
O\,{\sc ii}      &  8.76 $\pm$ 0.27 (28)   &  $-$0.06 $\pm$ 0.20 (23)  & ~~~   & 9.06 $\pm$ 0.44 (26) & +0.17 $\pm$ 0.35 (19)      \\
Mg\,{\sc ii}     &  7.50~~~~~~~~~~~(1)     &  +0.28~~~~~~~~~~~(1)      & ~~~   & 7.70~~~~~~~~~~~~(1)  & +0.48~~~~~~~~~~~(1)    \\
Al\,{\sc iii}    &  6.35 $\pm$ 0.04 (2)    &  +0.18~~~~~~~~~~~(1)      & ~~~   & 6.65 $\pm$ 0.03 (2)  & +0.47~~~~~~~~~~~(1)    \\
Si\,{\sc ii}     &  6.90~~~~~~~~~~~(1)     &  +0.23~~~~~~~~~~~(1)      & ~~~   & 6.94 $\pm$ 0.12 (4)  & +0.44 $\pm$ 0.07 (4)   \\
Si\,{\sc iii}    &  7.58 $\pm$ 0.12 (5)    &  +0.28 $\pm$ 0.26 (5)     & ~~~   & 8.00 $\pm$ 0.11 (3)  & +0.58 $\pm$ 0.18 (3)    \\
S\,{\sc ii}      &  7.17~~~~~~~~~~~(1)     &  +0.12~~~~~~~~~~~(1)      & ~~~   & 7.51~~~~~~~~~~~~(1)  & +0.33~~~~~~~~~~~(1)    \\
Ar\,{\sc ii}     &  6.59~~~~~~~~~~~(1)     &  +0.09~~~~~~~~~~~(1)      & ~~~   & --                   & --    \\
Fe\,{\sc iii}    &  6.45~~~~~~~~~~~(1)     &  $-$0.18~~~~~~~~~~~(1)    & ~~~   & --                   &  --    \\
\hline
\end{tabular}
\end{table*}

\subsection{LS5130}
\label{ls5130_dis} 
The value of microturbulence adopted for LS5130 and
$\gamma$\,Peg plays an important role in the derived abundances of
oxygen, magnesium and silicon in particular. As the abundance
variations of these elements are of particular interest, 
a short discussion is warranted. From the O\,{\sc ii} and
N\,{\sc ii} lines of $\gamma$\,Peg, the microturbulence
is likely to lie in the range of 5 $\pm$ 2 kms$^{-1}$. This is
consistent with other studies (e.g. Peters 1976, Gies \& Lambert 1992)
who quote a value at the lower end of this scale (3\,kms$^{-1}$) and
Ryans et al. (1996) who estimate a value of 6 $\pm$ 5 kms$^{-1}$. It is
likely that the latter error bars are too conservative, and from our
investigation we believe the result appropriate to our methods to be
in the range 3--7\,kms$^{-1}$.  For LS5130 we similarly derive a 
value of $5 \pm3$\kms, which leads to a relatively normal O abundance 
in this star, but significantly higher Mg and Si abundances. 
The latter
differential abundances would move closer to the value of O (as would
generally be expected from a chemical evolution point of view ; see
discussion in Section\,\ref{discussion_ii}), if the value of $\xi$ 
was increased in LS5130 and decreased in $\gamma$\,Peg by about 
3\,kms$^{-1}$ in each case. However while such changes are compatible 
with our random error range quoted, it would be an artificially 
biased analysis.  Indeed our best estimates of $\xi$ for
LS5130 tend to suggest that a value of {\em no more} than
5\,kms$^{-1}$ is required to flatten the equivalent width -- abundance
relation for the O\,{\sc ii} and N\,{\sc ii} species.   
One should note that the large number of O\,{\sc ii}
features (a total of 43), which renders this element abundance 
estimate particularly reliable. The analysis here implies that its
abundance is very similar to that of $\gamma$\,Peg, i.e.  no
significant enhancement of this element is observed, and this is a
very robust result. The other elements which we believe give reliable
results are C, N, Mg, Si and S -- and we note that all of these
elements show enhanced patterns, particularly Si and Mg. Although the
Mg abundance is based on only one line (the doublet at 4481\AA), it is
a strong, well measured feature which is reliably modelled in B-type stars
at this effective temperature. This pattern of a relatively normal oxygen
abundance together with enhanced C, N, Mg, Si and S is also reflected in
the other B-type stars. The values for
Al\,{\sc iii}, Ar\,{\sc ii}, and Fe\,{\sc iii} are based on relatively
weak detections of features and should be treated with some caution --
this is generally true for each of the four stars, and LS5130 has the
highest ``quality'' spectrum available, i.e. it has a high
signal-noise-ratio and the star itself has a relatively low projected
rotational velocity. 

\subsection{LS4419} 
\label{ls4419_dis} 
The higher projected rotational velocity of this star results
in fewer metal line features being identified.
However again the oxygen abundance is well constrained,
with 28 features contributing to the final result. Hence, we
conclude that LS4419 also shows no evidence for enhanced oxygen
(with respect to the solar neighbourhood), and
there is certainly no evidence for the large enhancements one might expect
to see in inner Galactic regions.  It should be noted (and this is
true for LS4784 and LS5130 also) that although the {\em absolute} Mg
abundance derived is close to the `solar' range of $7.50 \pm
0.2$\,dex, the value derived for $\gamma$\,Peg by our
methods (and indeed any other
star of approximately the same spectral type in the solar
neighbourhood) is 7.2\,dex. We believe that a value of 7.5
is $+0.3$\,dex {\em above} the solar neighbourhood value, and that the
absolute abundance discrepancy is possibly due to atomic data uncertainties
or non-LTE effects. Again we believe that the C, N, O, Mg, Si, and S 
differential abundances are reliable for this star, but the others 
should be treated with caution due to the paucity and weakness of 
the individual features.

\subsection{LS4784} 
\label{ls4784_dis} 
The quality of this spectrum is similar to that of
LS4419, and the previous comments about the reliability 
of the abundances (particularly with reference to O and Mg)
are equally applicable here. One should also note
that although the absolute abundances derived from the Si\,{\sc ii}
lines are of dubious worth (as discussed above with respect to $T_{\rm
eff}$ determinations), in both the standard stars and the LS stars, it
is encouraging to see that the {\em differential} abundances derived
from Si\,{\sc ii} and Si\,{\sc iii} are consistent. This suggests that
the same effect occurs in each spectra, and that the lines can be
quantitatively compared from star to star. 
Again in this star we see a relatively
normal oxygen abundance, but enhanced magnesium and silicon. Nitrogen
and sulphur are also enhanced, with carbon showing a mildly enriched
abundance.

\begin{table*}
\caption[]{Absolute elemental abundances for the star LS5127, together with a 
line-by-line differential analysis with the ``standard'' star $\iota$~Her
(HR6588) which is chemically ``normal'' with very similar atmospheric
parameters.  
}
\label{LS5127_abundances}
\vspace{0.25cm}
\begin{tabular}{llll}\hline
Ion          &       LS5127           & $\iota$\,Her        & Differential \\
\hline 
C\,{\sc ii}      &  8.82 $\pm$ 0.34 (2)  &   8.26 $\pm$ 0.27 (11)  & +0.55~~~~~~~~~~~(1)  \\
N\,{\sc ii}      &  8.29 $\pm$ 0.10 (7)  &   7.87 $\pm$ 0.18 (23)  & +0.51 $\pm$ 0.23 (6)  \\
O\,{\sc ii}      &  9.05 $\pm$ 0.19 (7)  &   8.79 $\pm$ 0.20 (54)  & +0.21 $\pm$ 0.25 (6)     \\
Mg\,{\sc ii}     &  7.76~~~~~~~~~~~(1)   &   7.23 $\pm$ 0.26 (6)   & +0.69~~~~~~~~~~~(1)    \\
Al\,{\sc iii}    &  6.68~~~~~~~~~~~(1)   &   6.23 $\pm$ 0.05 (6)   & +0.36~~~~~~~~~~~(1)  \\
Si\,{\sc ii}     &  7.45 $\pm$ 0.26 (3)  &   6.84 $\pm$ 0.50 (4)   & +0.81 $\pm$ 0.33 (3)   \\
Si\,{\sc iii}    &  8.18 $\pm$ 0.12 (4)  &   7.51 $\pm$ 0.20 (7)   & +0.64 $\pm$ 0.08 (3)  \\
S\,{\sc ii}      &  7.44 $\pm$ 0.16 (5)  &   6.99 $\pm$ 0.23 (41)  & +0.39 $\pm$ 0.49 (5) \\
S\,{\sc iii}     &  7.97 $\pm$ 0.34 (2)  &   7.14 $\pm$ 0.30 (5)   & +0.76 $\pm$ 0.02 (2)  \\
Fe\,{\sc iii}    &  --~~~~~~~~~~~~~~~(0)  &   6.97 $\pm$ 0.67 (8)   & --~~~~~~~~~~~~~~~~~(0)  \\
\hline
\end{tabular}
\end{table*}

\begin{table*}
\begin{flushleft}

\caption[]{The errors listed in this table illustrate how the
absolute abundances in the B-type stars would vary within the error
bars of the atmospheric parameters. A {\em systematic} error in {\em
both} the target LS-star and the standard used in the differential
abundance analysis would result in little change in the differential
results. The results below are for
typical random analysis errors of $\log g \pm0.2$, $T_{\rm eff}
\pm1000$\,K and $\xi \pm3$\,kms$^{-1}$. The values were calculated
around a model atmosphere with $\log g = 3.6, T_{\rm eff} = 21000, \xi
= 5$.}

\label{typical_errors}
\vspace{0.25cm}
\begin{tabular}{lllllll} \hline
 Species  & $\log g+0.2$ & $\log g-0.2$ & $T_{eff}$+1000 & $T_{eff}-$1000 & $\xi$+3km/s & $\xi-$3km/s \\\hline

 C\,{\sc ii}   &      \,\,\,\,0.01  &  \,\,\,\,0.00  &  \,\,\,\,0.02 &   \,\,\,\,0.03  &  $-$0.08  &  \,\,\,\,0.11  \\
 N\,{\sc ii}   &      \,\,\,\,0.07  &  $-$0.05  &  $-$0.06 &   \,\,\,\,0.12  &  $-$0.08  &  \,\,\,\,0.11  \\
 O\,{\sc ii}   &      \,\,\,\,0.12  &  $-$0.11  &  $-$0.18 &   \,\,\,\,0.23  &  $-$0.11  &  \,\,\,\,0.15  \\
 Mg\,{\sc ii}  &      $-$0.05  &  \,\,\,\,0.05  &  \,\,\,\,0.13 &   $-$0.11  &  $-$0.22  &  \,\,\,\,0.32  \\
 Al\,{\sc iii} &      \,\,\,\,0.05  &  $-$0.03  &  $-$0.02 &   \,\,\,\,0.08  &  $-$0.08  &  \,\,\,\,0.15 \\
 Si\,{\sc ii}  &      $-$0.06  &  \,\,\,\,0.08  &  \,\,\,\,0.20 &   $-$0.15  &  $-$0.05  &  \,\,\,\,0.11 \\
 Si\,{\sc iii} &      \,\,\,\,0.09  &  $-$0.09  &  $-$0.14 &   \,\,\,\,0.18  &  $-$0.21  &  \,\,\,\,0.24 \\
 S\,{\sc ii}   &      $-$0.04  &  \,\,\,\,0.05  &  \,\,\,\,0.16 &   $-$0.14  &  $-$0.09  &  \,\,\,\,0.21 \\
 S\,{\sc iii}  &      \,\,\,\,0.13  &  $-$0.12  &  $-$0.14 &   \,\,\,\,0.18  &  $-$0.07  &  \,\,\,\,0.13 \\
 Ar\,{\sc ii}  &      \,\,\,\,0.01  &  \,\,\,\,0.02  &  \,\,\,\,0.12 &   $-$0.06  &  $-$0.06  &  \,\,\,\,0.14 \\
 Fe\,{\sc iii} &      \,\,\,\,0.09  &  $-$0.08  &  $-$0.04 &   \,\,\,\,0.08  &  $-$0.08  &  \,\,\,\,0.24 \\
\hline
\end{tabular}  
\end{flushleft}
\end{table*}

\subsection{LS5127} 
\label{ls5127_dis} 
This star is cooler than the three previously discussed, and
hence we used a different solar neighbourhood star for the
differential abundance comparison.  The star $\iota$\,Her has
atmospheric parameters more compatible with that of LS5127 than
$\gamma$\,Peg (we did in fact carry out the same differential
analysis using $\gamma$\,Peg, and the results were not significantly 
different to
those in Table\,~\ref{LS5127_abundances}). The oxygen abundance appears
to be slightly above normal, but is not as high as the abundances of
both Mg and Si. We find a particularly large magnesium abundance from
the 4481\AA~ feature, and a line profile together with a spectral
synthesis of the region is shown in Fig.\,1. The microturbulence
derivation in LS5127 is probably more uncertain than in the previous
three stars, due to the fewer O\,{\sc ii} lines visible at this
effective temperature. However, it is unlikely that the particularly large
Mg\,{\sc ii} abundance can be explained by an underestimate of
$\xi$. In Fig.\,1 we have used $\xi=7$\,kms$^{-1}$, but adopting a
value of 10\,kms$^{-1}$, and a high abundance for Al\,{\sc iii}
(i.e. $>$7.00 in order to ensure that the strength of this feature is
not due to severe blending with the Al~{\sc iii} line at 4479\AA) 
still results in the
differential value being $>$0.5\,dex. Again we find that both
ionization stages of Si give consistently high abundances, consistent
with the Mg enhancement. 

\begin{figure}
\psfig{file=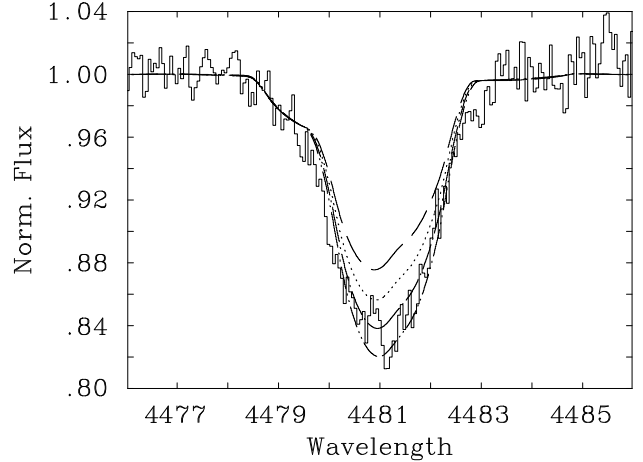}
\label{LS5127_mgii_line}
\caption[]{The Mg\,{\sc ii} doublet at 4481\AA~ in LS5127 with
synthesised fits.  The observed line profile is the solid histogram,  
together with theoretical fits for magnesium abundances of
7.2, 7.5, 7.8 and 8.1\,dex (each with a microturbulence of $\xi$=7
kms$^{-1}$), convolved with a rotational broadening function $v\sin i
= 120$\,kms$^{-1}$.  A value of 7.8\,dex is required to fit the line
properly.  Note that there is a Al\,{\sc iii} line in the blueward
wing, and we have also assumed a high Al abundance (6.8 dex)when 
synthesizing this region.}
\end{figure}

\section{Distances, Galactic positions and evolutionary status}
\label{dist_discuss}
We have calculated distances to the stars as follows.  Masses
were derived by placing the stars in the effective temperature -
surface gravity diagram using the evolutionary tracks of Schaller et
al. (\cite{Sch92}).  This allowed a calculation of the luminosity
appropriate to the stars {\em current} evolutionary status, 
rather than assuming that the object is on the zero-age
main sequence (ZAMS).  The absolute visual magnitude was then
determined employing the bolometric corrections of Kurucz
(\cite{Kur79}). Extinctions were estimated from the measured
$B-V$ values taken from Reed (\cite{reed93}), and the intrinsic colours of
Deutschman et al. (\cite{Deu76}) for stars with these atmospheric
parameters.  The distance was finally calculated assuming a standard
Galactic extinction law, i.e. with $R=3.1$ (Seaton \cite{Sea79}).
We note that although our stars have relatively low $v\sin i$ 
values, their rotational velocities could be higher depending on the
inclination viewing angle. Recent studies of the evolution of
rotating stars (e.g. Heger \& Langer \cite{heg2000}, Maeder \& Meynet
\cite{maed2000}) indicate that their main-sequence properties
maybe altered by rotation, however the differences predicted in the
luminosities are smaller than the uncertainities in the rest of the
distance calculation and do not significantly increase the aerror 
bar we calculate on the stellar distances. 

The four B-type targets lie at significant distances below the plane
of the Galaxy (see Table~\ref{distance_data}), and it is unlikely that
they have formed at their current positions. The probable explanation
is that they were born in the Galactic disk, and were then
subsequently ejected. This class of early-type stars lying at, in some
cases, significant distances from their formation sites (i.e. runaway
stars) has been studied extensively, and their existence is well
accepted (see Rolleston et al. \cite{roll99}).   
In a previous paper (Smartt et al. \cite{sdl97}) we have discussed
the analysis of four blue supergiants lying at similar Galactic
latitudes, and a full discussion of the possible origins and ejection
mechanisms from the disk of these types of objects is
presented. Leonard (\cite{Leo93}) has presented plausible mechanisms
for early-type stars suffering large ejection velocities within
massive binary evolution scenarios, or dynamical ejection from
clusters. These can result in initial velocity injections of up to
$\sim$ 350 kms$^{-1}$, easily allowing our objects to reach their
current estimated distance from the Galactic plane (z).

\begin{table*}
\caption[]{Distances, masses, lifetimes, radial velocities and regions of 
origin for the programme stars. Masses and evolutionary lifetimes were 
determined from the stellar evolutionary tracks of Schaller et al. (1992).}
\vspace{0.25cm}
\label{distance_data}
\begin{tabular}{rccccccc}\hline
Star     & M  & T$_{ev}$ & d$_{\odot}$ & $z$ & PRV & V$_{i}$ & R$_{g}$ \\
         & (M$_\odot$) & (Myrs) & (kpc) & (kpc) & (kms$^{-1}$) & (kms$^{-1}$) & (kpc)\\\hline
LS5130 & 10$\pm$2 & 24$\pm5$ & 6.9$\pm$1.0 &  0.7 & $-$7  & 64 & 3.1$\pm$1.0 \\
LS4419 & 11$\pm$2 & 20$\pm$5 & 5.2$\pm$1.0 &  0.5 & $-$28 & 45 & 2.9$\pm$1.0 \\
LS4784 &  8$\pm$2 & 27$\pm$5 & 3.8$\pm$1.0 &  0.4 & $-$30 & 36 & 4.7$\pm$1.0 \\
LS5127 & 10$\pm$2 & 25$\pm$5 & 9.1$\pm$1.0 &  1.2 & $-$3  & 98 & 2.5$\pm$1.0 \\
\hline 
\end{tabular}
\end{table*}

In order to constrain the regions of the disk where these stars were
formed, we follow the arguments of Smartt et al. (\cite{sdl97}). For each
of the stellar spectra we have determined a radial velocity, and we
compare this velocity to that expected if the stars were co-rotating
with the disk at their current projected positions (assuming a flat
rotation curve and R$_{\odot}$ = 8.5\,kpc, $\Theta = 220$\,\kms from
Kerr \& Lynden-Bell \cite{ker86}).  We calculate a peculiar radial
velocity (PRV) which is the difference between these two, as a
representative measure of the dynamics of the stars with respect to
the Galaxy rotation. This PRV is only one component of the total
peculiar space velocity, with the other two components unknown
(directed along axes mutually perpendicular to the PRV). We assume
that the second component of velocity is perpendicular to the Galactic
plane (which, given the small Galactic latitudes of these stars is a
reasonable assumption). House \& Kilkenny (\cite{hou80}) have given an
expression for the force field normal to the Galactic plane ($K_{\rm z}$), 
but this is probably only valid for $R_{\rm g} \sim 10$\,kpc. Given 
that these stars have origins closer to the Galactic Centre, we have assumed 
a simple relation which is an approximation to the shape of
the House \& Kilkenny form of $K_{\rm z}$, but which is always greater
in magnitude (with a peak of twice the House \& Kilkenny maximum). 
We assume that within $0 \leq z \leq 700$\,pc the force field
$K_{\rm z}$ is approximated by $-kz$ (with 
$k = 2.29\times10^{-11}$\,cms$^{-2}$pc$^{-1}$) and for $z > 700$\,pc then 
$K_{\rm z} = -k$. Analytical expressions for 
the initial velocity (V$_{i}$) and time required for the 
stars to reach their present $z$ distance, could then be deduced.
The initial ejection velocity perpendicular to the Galactic plane 
needs to be no more than 100\,\kms (see Table\,\ref{distance_data})
for these stars to reach their
present positions, and significantly less for three of the objects.
Velocities of this order are well within those plausible for 
common ejection methods (Leonard \cite{Leo93}). The PRV of each star
is also relatively low, indicating that the initial space 
velocities of these stars were probably directed mostly in the $z$-direction.
However without proper motion information we cannot definitively constrain
their space velocity. If we assume that the 
mean ejection velocity in the sample (V$_{i} \sim$ 60\kms)
is indicative of the maximum value of any one velocity component, then 
we can calculate a circular locus within which the star probably 
formed. In 20\,Myrs, a star could have moved 1.7\,kpc within 
(or parallel to) the Galactic disk, and this is a reasonable 
error on the birth sites of the stars. 

\begin{figure}
\psfig{file=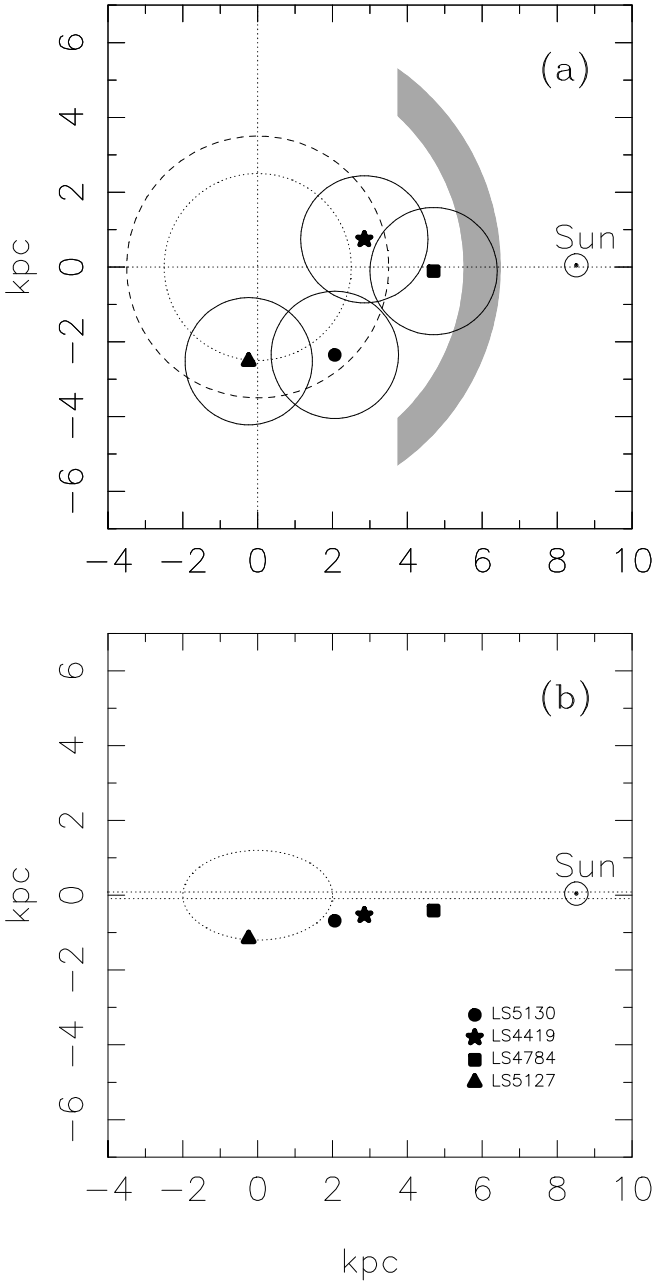}
\label{galactic_pos}
\caption[]{{\bf (a):} A schematic view of the Galaxy with the
Galactic Centre at (0, 0). 
The inner dotted circle represents the possible bulge-disk 
transition with which a rapid decline in the {\em disk} stellar
density is associated; the outer dashed circle represents the 
stellar ring proposed by Bertelli et al (1995) of radius 3.5\,kpc
(they estimate a Gaussian half-width of $\sigma_{ring}=0.5$\,kpc). 
The shaded arc is a schematic representation
of the Sagittarius spiral arm at R${_g}=5.5$\,kpc. Also shown
are possible birth sites in the disk, together with our error estimates. 
{\bf (b):} A section through the Galaxy with the current positions of the
stars marked, again the Galactic Centre is at (0, 0). The disk scale height
for early-type stars is shown (~90\,pc), together with the
spheroidal latitude extent assuming $a/b=0.8$ (Bahcall 1986).
}
\end{figure}

Figure~\ref{galactic_pos} illustrates the estimated stellar birth sites
and also schematically draws attention to features of 
Galactic structure in this region.
Bertelli et al. (\cite{Bert95}) have discussed the possibility that 
a `stellar ring' exists, which surrounds the Galactic Centre at a
distance of 3.5\,kpc, with a 0.5\,kpc half-width. 
The nature of this ring is still somewhat uncertain, but Bertelli et al. 
claim that assuming it to be a region of increased stellar disk density 
would produce synthetic colour-magnitude diagrams closely matching those
they have observed in fields towards the Galactic Centre. They suggest 
that it is associated with active star formation, and that 
the cross-section through this ring is at a maximum when viewed at 
$|l|=23.5^o$. They further suggest that this is consistent with an
observed peak in OH maser sources (Blommaert
et al. \cite{blomm94}) around $l=25^o$, a signature of 
relatively recent star formation. Whether this is a complete ring 
surrounding the inner Galaxy, or part of an inner spiral arm, 
is still debatable (Ortiz \& L\'{e}pine \cite{ort93}).  
Clemens et al. (\cite{cle88}) have mapped the distribution
of molecular gas in the first quadrant of the Galaxy, and found an H$_{2}$
ring at approximately 4\,kpc from the Centre. This would appear to 
correlate with the stellar ring found by Bertelli et al.,  but the 
latter suggest that the stellar and molecular ring are not necessarily 
co-incident. Rather that the molecular ring maybe 
the result of a ``bow-shock of propagating star formation''. 

Although some Galaxy models (Bahcall \& Soneira 1980; Bahcall 1986) adopt 
a relation for the number density of disk stars
which falls off exponentially with 
increasing distance from the Galactic Centre, 
it appears that at  R$_g\leq$2.5\,kpc
from the Galactic Centre there is a sharp decrease in the stellar
disk density (Bertelli et al 1995, Paczynski et al. \cite{pacz94}). 
This has been interpreted as the transition between the disk and the bulge. 
Figure~\ref{galactic_pos} suggests that the four B-type stars 
have their origins in this stellar ring, or where the molecular ring
dynamics are causing current star formation. Indeed if there is a 
disk/bulge transition, early-type stars may be relatively rare within
this inner region, although current and
spectacular star formation is certainly on-going in the inner few
hundred parsecs (Serabyn et al. 1998; Najarro et al. 1997). The above 
indicates that our B-type stars probably formed around R$_g\sim$3.5\,kpc
in the Galactic disk and were subsequently ejected.

Carr et al. (\cite{carr2000}) have presented 
IR spectra of a red M-type supergiant (IRS7) lying 
in the central cluster, and compared the metal line spectra with 
that of a similar star in the solar neighbourhood. A differential
abundance analysis of the star indicates that it has an Fe 
abundance very close to solar, and certainly does not appear
metal enhanced in any significant way (although it does appear to 
show CNO-abundances typical of severe dredge-up after core H-burning). 
At first this may seem a surprising
result given that one might expect a linear Galactic abundance gradient 
in the thin disk to keep rising towards the Centre, and that the central
regions would thus show super-metal rich abundances. However 
there is lack of molecular gas, young stars and H\,{\sc ii} regions
within the inner 3.5\,kpc, until one gets to the 
central cluster. Hence, whatever the evolutionary history of the 
massive stellar clusters at the Centre, it seems to have occurred
independently of the Milky Way disk ($R_{\rm g} \gtrsim 3.5$\,kpc). 
There is no reason therefore too assume that the disk and central cluster
are chemically linked, and one should consider them as separate 
entities in evolutionary scenarios. 

\section{Discussion}

\subsection{The oxygen abundance gradient towards the Galactic Centre}
\label{discussion_i}
\begin{figure}
\psfig{file=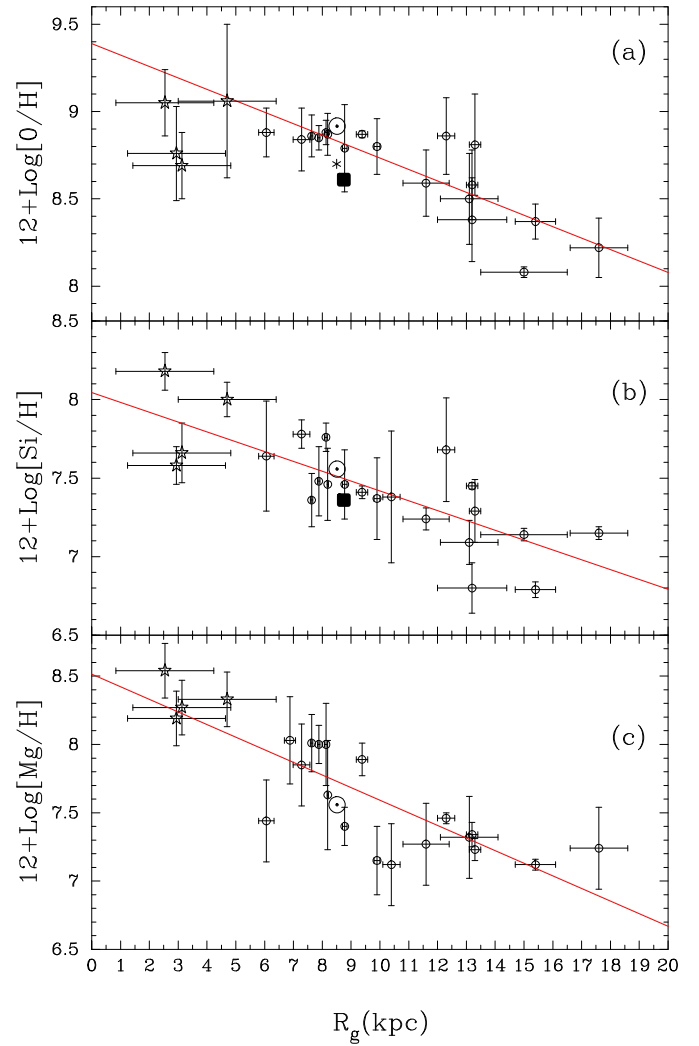}
\caption[]{
Galactic abundance gradients and the stellar sample (open circles)
of RSDR. Also shown are the four current targets (asterisks) at 
$R_{\rm g} < 5$\,kpc, the error bars being the 
standard deviations in the mean line abundances or
conservative $\pm$0.2\,dex for Mg.  In 
{\bf (a)} the abundance gradient plotted is a least squares fit to the
data in RSDR {\em only} and has a value of
$-$0.07\,dex\,kpc$^{-1}$; the four stars towards the Galactic Centre
clearly lie below this gradient if it were extrapolated towards the
centre. In {\bf (b)} and {\bf (c)} the linear least squares gradients are
fits to {\em all} the points in the plot. 
The Si and Mg abundances in the Galactic Centre{\em
are} compatible with a steadily increasing abundance gradient towards
the centre, and a re-evaluation of the least-squares fit through 
all points leads to values of $-0.06 \pm0.01$\,dex\,kpc$^{-1}$ for
Si and  $-0.09 \pm0.02$\,dex\,kpc$^{-1}$ for Mg. 
The solid square is the Orion abundance from Cuhna \&
Lambert (1994), the asterisk is the Local ISM from Meyer et
al. (1998), and the solar symbol is the solar abundance from Grevesse
\& Noels (\cite{gre93}).}
\label{OSiMg_gradients}
\end{figure}

As discussed in Sections\,\ref{ls5130_dis}--\ref{ls5127_dis}, all
four stars have relatively normal oxygen abundances. 
If the oxygen abundance gradients derived in the solar neighbourhood and 
outer disk are extrapolated inwards
(Smartt \& Rolleston 1997, RSDR; $-$0.07 dex\,kpc$^{-1}$) their O abundances 
would be 0.3$-$0.4\,dex above the solar neighbourhood value. 
LS5130 and LS4419 show no evidence at all of oxygen-rich photospheres,
and both sample the O\,{\sc ii} line spectra extremely well, with
43 and 28 features measured respectively. 
If the statistical errors are normally distributed around the mean,
then the random error in our mean differential
abundance ($ \sigma/\sqrt n $; where $\sigma$ is the standard
deviation quoted in Tables\,\ref{LS5130_abundances} and 
\ref{LS4419_abundances}) may be as small as $\pm$0.04 for both these stars. 
Furthermore, a systematic error in the atmospheric 
parameters significantly greater than that quoted in 
Table\,\ref{typical_errors} would be required to make these stars 
O-rich which would have major consequences for the other elements
and for the continuum fitting of the $uvby$ indices. The large number of 
O\,{\sc ii} lines well observed in these stars and the 
possible error sources in Table\,\ref{typical_errors} make this a 
robust result. The other two stars LS4784 and LS5127 show
marginal evidence for oxygen abundances higher than their
comparison stars, with standard errors of the mean $\pm$0.1\,dex. 
However again these results appear to be incompatible with
an O abundance that is 0.3$-$0.4\,dex above solar. 
A number of previous authors have speculated that the 
oxygen abundance gradient (in particular) may steepen towards the Centre 
(V\'ilchez \& Esteban \cite{vil96}, Shaver et al. \cite{Sha83}). 
However, we can find no evidence that this is the case. 
Figure\,\ref{OSiMg_gradients} shows the oxygen abundances of our stars
together with those of the large Galactic B-star sample of RSDR. 
The low metallicity stars in the outer Galaxy ($R_{\rm g} > 11$\,kpc) 
are clearly evident, but between 6$-$9\,kpc there is little evidence
of O abundance variations. The four new stellar points toward
the Centre accentuate this relation, and it would appear that the O
abundance gradient in the Galaxy flattens off at around 9\,kpc and 
does not significantly increase at smaller Galactocentric distances. 

\begin{figure}
\psfig{file=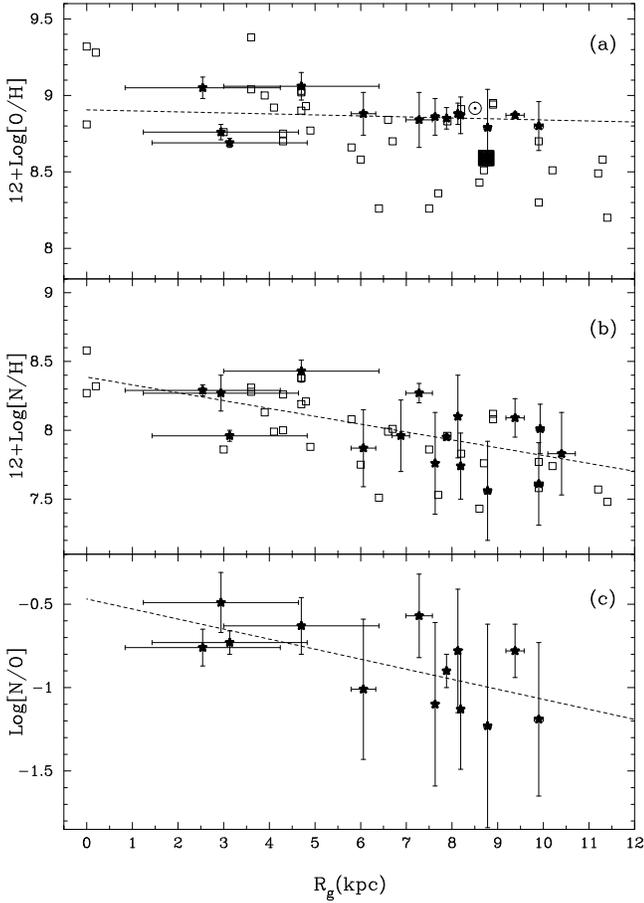,width=9cm}
\caption[]{
{\bf (a):} Filled stellar symbols are oxygen
abundances of the Galactic Centre objects, together with stars 
from RSDR with $R_{\rm g} < 11$\,kpc. The 
error bars for the Galactic Centre stars are standard errors 
of the mean. The dotted line is a least-squares straight-line fit
to all the B-stars. It has a gradient of $-0.01 \pm0.01$ dex\,kpc$^{-1}$.
The open symbols are results from H\,{\sc ii} regions from Afflerbach
et al. (1997). Solar and orion points as in Fig\,3. 
{\bf (b):} Symbols are as in (a), but for nitrogen abundances. The 
gradient derived is $-0.06 \pm 0.02$\,dex\,kpc$^{-1}$, on good agreement
with the outer gradient as derived in RSDR. Reasonable agreement
between the stellar and H\,{\sc ii} region points is found. 
{\bf (b):} The $\log$N/O ratio, from the stars in the above two panels. 
The straight-line fit is $-0.06\pm0.02$\,\dk, indicating a fairly 
significant trend in N/O with Galactocentric distance. }
\label{ON_HII_comp}
\end{figure}

Afflerbach et al. (\cite{Aff97}) have carried out an extensive
study of the N, S and O abundances in H\,{\sc ii} regions  with
$R_{\rm g}\leq 12$\,kpc. In Fig.\,\ref{ON_HII_comp}, their O 
abundances are plotted together with those for our stars 
and the stellar sample of RSDR with $R_{\rm g}\leq 12$\,kpc.
The results from the H\,{\sc ii}
regions and the RSDR data-set in the outer Galaxy are 
in excellent agreement (see RSDR and Smartt \cite{sma2000} for 
further details). However differences appear at smaller 
Galactocentric distances. In particular while the
H\,{\sc ii} region data is consistent with a linear gradient
$-$0.06\,dex\,kpc$^{-1}$, no significant variations are apparant
for the B-stars. One possible explanation is that the zero-points 
for the two studies is somewhat different. For example,
if the Afflerbach et al. 
data were  shifted upwards by $\sim$0.2\,dex (the difference in the
mean results within $7.5 < R_{\rm g} < 9.5$kpc), then
good agreement would be found within the solar neighbourhood, 
but the B-stars towards the Centre would not reflect the 
H\,{\sc ii} region results. The cause of this discrepancy is unclear
and appears to be limited to oxygen (see Sect.\ \ref{discussion_ii}
and  \ref{discussion_iii}). We note that Afflerbach et al. 
is the only H\,{\sc ii} region study of the inner Galaxy to quote
oxygen abundances. Rudolph et al. (1997) were unable to 
independently determine S and O, while Shaver et al. (1983) point out
that their two inner Galaxy objects ($R_{\rm g}\sim$ 4\,kpc) 
are somewhat peculiar, and that their radio and optical sources
may not be the same. Afflerbach et al.\ have
recalculated N abundances from the data of Simpson et al. (\cite{sim95}) 
and find differences of a factor of $\sim$2
in some of the highest 
metallicity regions. Additionally,
on average there seems to be a 50\% discrepancy in the sulphur
abundances. Therefore it appears 
that the nebular abundances in the inner Galaxy require further 
study to constrain their absolute values, while  a
more extensive sample of B-stars is required. Smartt et al. 
(\cite{smt2000}) have discussed the problem of determining
reliable absolute nebular abundances in regions of solar
abundance and above in M31, and find significant discrepancies
across different analysis methods.

\begin{figure}
\psfig{file=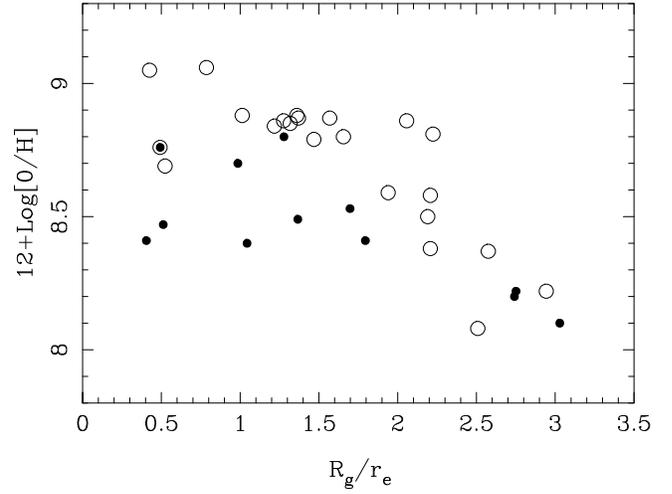,width=9cm}
\caption[]{Oxygen abundances in the Milky Way from our stellar
sample (open circles) and the H\,{\sc ii} region 
oxygen abundances in NGC2403 from Garnett et al. (1997, solid
circles). Both are shown as a function of effective radius. 
Within $R_{\rm g}/r_{\rm e} \lesssim 1.8$ there is little
evidence for increasing oxygen abundances in either galaxy, but 
there are many other well studied spirals which do not show such 
an effect.}
\label{mw_ngc2403_grads}
\end{figure}

In terms of effective radius (where $r_{\rm e}=5.98$\,kpc, from de 
Vaucouleurs \& Pence \cite{dev78}), the break in the observed O abundance
gradient occurs at $R_{\rm g}/r_{\rm e} \simeq1.8$. The spiral galaxy
NGC2403 has been studied by Garnett et al. (\cite{gar97}) who estimated 
H\,{\sc ii} region abundances at 12 points across the disk. 
In Fig\,\ref{mw_ngc2403_grads} the oxygen abundance is plotted
as a function of effective radius for both these galaxies.
A similar gradient change may be present in NGC2403, although 
the sampling is relatively sparse. 
This would imply that a two component model for the 
radial variation of oxygen is not exclusive
to our own Galaxy and suggests indirect evidence that it is a 
real effect. Certainly the data suggests that for both 
galaxies, at  $R_{\rm g}/r_{\rm e} \lesssim 1.8$, there is little
evidence for increasing oxygen abundances.
However, the well studied  galaxy M33 (Garnett et al.
\cite{gar97}, Henry  \& Howard \cite{hen95}, Monteverde et al.
\cite{mon2000}) which is similar in size, mass and metallicity 
to NGC2403 does not appear to show such a break. Additionally,
M101 and NGC4303 are other examples of extensively studied galaxies
(Kennicutt \& Garnett \cite{ken96}, Martin \& Roy \cite{mart92}), 
with 30$-$80 H\,{\sc ii}
region data points sampled in each. Neither show evidence
for anything other than a linear relation across the whole
disk.

\subsection{Magnesium, Aluminium, Silicon and Sulphur abundances 
towards the Galactic Centre}

\label{discussion_ii}

The magnesium and silicon abundances for the Galactic 
Centre stars and the 
corresponding RSDR dataset are plotted in Fig\,\ref{OSiMg_gradients}. 
For oxygen and silicon we note that the absolute abundances derived in our
two solar neighbourhood standard stars are similar to the mean of 
the RSDR data set within 1\,kpc of the Sun. Hence the 
absolute results of the Galactic Centre stars are on a consistent scale to 
the rest of the data. However with Mg, 
the abundance derived in $\gamma$~Peg and $\iota$~Her
is 7.2\,dex, whereas in RSDR the mean local value is 7.85 dex
(again estimated from the mean of stellar values 
within 1\,kpc of the Sun).
The stars analysed in this paper have effective temperatures 
in the range $17700-22100$\,K ,while the RSDR sample have a mean
temperature of 25275\,K and it is possible that the modelling of this line is
temperature dependent. It is important to use homogeneous datasets when 
looking for abundance trends and that is indeed what was done in RSDR. 
They showed that although the zero-point of any particular
abundance may be in the error, any gradient biasing is removed by using 
suitable sub-samples of stars. Hence to put the four Galactic Centre stars 
onto the same scale as RSDR, we have added 7.85\,dex to
their {\em differential} Mg abundances. 
This assumes that $\gamma$~Peg, $\iota$~Her and the stars within 1\,kpc
of the Sun in RSDR have similar abundances and that any offset is 
due to temperature dependent modelling inaccuracies. It appears
that the Galactic Centre stars are richer in Mg than
their local counterparts (see also Sect\,\ref{LS5127_mgii_line}). 
From Table\,\ref{typical_errors} an underestimate of the microturbulence
($\xi$) in each star could reduce this Mg overabundance as it is based on
a relatively strong line. However an increase in $\xi$ by 5\,\kms or more 
in each star would be required. While observational uncertainties
might cause a particular determination of $\xi$  to have such an error, 
there is no evidence that we have {\em systematically} underestimated 
the microturbulence in each star. The Mg abundance
gradient in Fig.~\ref{OSiMg_gradients} hence appears to steadily increase
towards the Centre. A value of $-0.09 \pm 0.02$\dk is derived by fitting
all points, which is slightly larger the RSDR result, although they 
agree within the uncertainties. 

\begin{figure}
\psfig{file=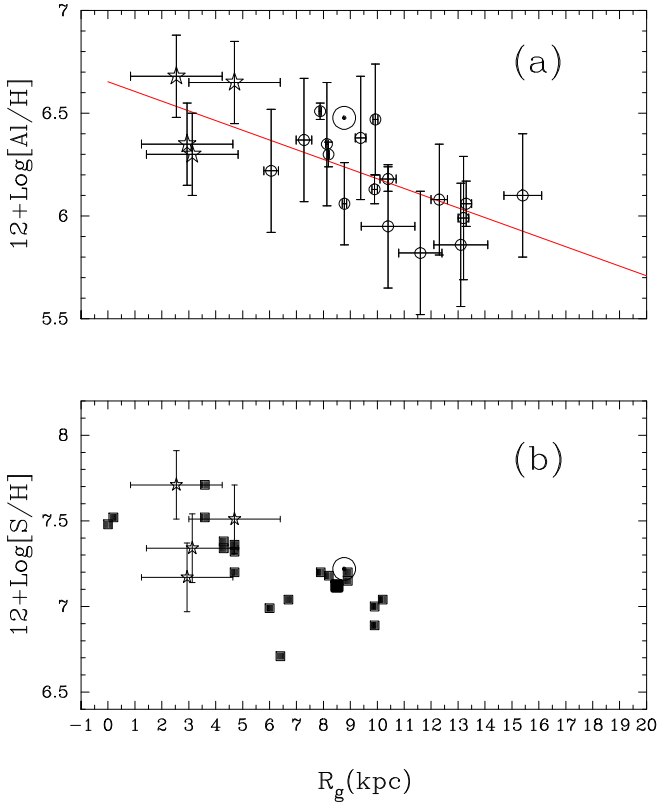,width=9cm}
\caption[]{
{\bf (a):} Asterisk symbols are aluminium
abundances of the Galactic Centre objects and circular points are
Al abundances from RSDR stars.  
Error bars for the Galactic Centre stars are taken as typically
$\pm$0.2 dex. The dotted line is a least-squares straight-line fit
for all the B-type stellar data, with a gradient of 
$-0.05 \pm0.01$ dex\,kpc$^{-1}$,
similar to that derived by RSDR.
The solar point comes from Table\,\ref{LS5130_abundances}
{\bf (b):} Asterisk symbols are stellar sulphur abundances
and filled squares are results from Afflerbach et al. (1997). Little
reliable stellar data exists for S in the outer Galaxy. }
\label{fig_SAl_grads}
\end{figure}

The silicon abundances of the Galactic centre stars are also plotted
in Fig.\ref{OSiMg_gradients}, and again appear compatible with 
a steadily increasing abundance gradient towards the centre. 
A fit through all the Si points gives a gradient of  $-0.06 \pm 0.01$\dk, 
again in good agreement with the RSDR value for the outer Galaxy only. 
The differential abundances derived for each star are between
0.3$-$0.6\,dex higher than their respective standards, strongly 
supporting the idea that these stars are richer in Si than 
solar neighbourhood material. 
In Fig.\,\ref{fig_SAl_grads} the Al abundances of the galactic 
Centre stars are plotted again with the data from RSDR.
Aluminium appears to increase steadily toward the Centre, and a 
gradient of $-0.05 \pm0.01$ dex\,kpc$^{-1}$ is derived from the whole
data set, similar to that produced by RSDR for the anti-centre direction.

Sulphur abundances for our stars are also available, although we
have no similar data at greater radii from RSDR. In 
Fig\,\ref{fig_SAl_grads} our stellar abundances 
are compared with with those from Afflerbach et al.
(1997). Reasonable agreement is found with
these stars appearing to be S-rich and having abundances similar to those
found in nebular studies towards the Centre (although the caveat 
regarding the nebular absolute values
discussed in the previous section may be important). 

The correlation of of O, Mg, and Si abundances
found in the outer Galaxy (see RSDR) and
in the low metallicity Magellanic Clouds (Rolleston et al. \cite{roll96})
is theoretically expected as these elements are all produced
and returned to the ISM through supernovae Type\,II. 
One would expect that this trend should continue towards the inner
Galaxy, and it is surprising that our O abundances show
no sign of increase, while Mg and Si appear significantly higher than 
their comparison stars. The S abundances in these stars are
also higher than normal, although the dataset is somewhat less robust
given the weakness of the absorption features of this element.

Prantzos et al. (\cite{pra94}) have investigated the 
effect of metallicity dependent
yields (taken from the model calculations of Maeder \cite{mae92}) on the 
galactic chemical evolution of C and O
and concluded that oxygen yields from massive 
stars tend to decrease towards higher metallicities. This is due to mass
loss in low metallicity stars, resulting in most of the star's mass being 
retained up until the final explosion. However, at high metallicities
significantly more mass is ejected in the stellar wind (which is rich in 
helium and carbon), and leaves a smaller mass fraction which can be converted 
into oxygen. The relatively high carbon and low oxygen 
abundances derived here are in 
qualitative agreement with the situation predicted by these models. 
A full study of the evolution of elements heavier than oxygen has not been 
undertaken using metallicity dependent yields. However Maeder (1992) has 
discussed how the combined yield of ``heavy elements'' (defined as Ne, Mg, Si,
S, Ca \& Fe) from massive stars varies with original metal composition.
Interestingly, he finds that the ``heavy element'' contributions of the 
stellar models at high metallicity ($Z=0.02$) are not as low as is the case
for oxygen. The situation is further complicated by the fact that chemical 
yields are dependent on the conditions for black hole formation since the 
final yield of a pre-supernova massive star (pictured in the conventional 
onion-skin model) depends on what fraction of the stellar layers are
retained during core collapse.
 
The observational findings presented here should provide a stimulus to 
investigations of the evolution of the Galaxy at high metallicities, given the 
probability that stellar yields are highly variable with metallicity, and 
dependent on the limits of black hole formation. Furthermore, we should attempt
to determine the true quantitative variation in the ratios of the 
$\alpha$-elements in the inner Galaxy to investigate if this trend is
real. The recent chemical evolution models of Portinari \& Chiosi 
(\cite{por99}, \cite{por2000}) do indicate that the O abundance gradient
may indeed flatten off in the region $3 \la R_{\rm g} \la 6$\,kpc,
which is encouraging. However their SN Type\,II yields would suggest that
the other $\alpha$-processed elements (Mg and Si) should follow a similar
trend, which we do not observe. Boissier \& Prantzos (\cite{bois99})
also suggest that their evolutionary models for the Milky Way predict
a flattening of the metallicity gradient with time, and that 
saturation is reached in the inner Galaxy. This is 
in reasonable quantitative agreement with what we find for oxygen 
abundances, but does not explain the behaviour of the other
elements.

\subsection{Carbon and nitrogen abundances towards the Galactic Centre}
\label{discussion_iii}

Each of the four stars do appear to have nitrogen abundances significantly
higher than their solar neighbourhood counterparts. The result for LS5130
is  more marginal than the others, but with 19 features measured an
error in the mean ($\sigma/\sqrt n$) of 0.03\,dex would suggest that 
the enrichment is real although small. Figure\,4(b) shows the nitrogen 
abundances of the Galactic Centre stars plotted along with a sample of 
stars from RSDR, and the H\,{\sc ii} regions from Afflerbach et al. (1997).  
The gradient found in RSDR appears to continue towards the Centre;
certainly the abundance estimates do not appear to flatten as for oxygen. 
Reasonably good agreement
with the absolute values of the H\,{\sc ii} region analyses is found, and
we calculate quite a significant gradient for $\log$\,N/O. The 
full sample of RSDR yielded a gradient in N/O of $-0.04\pm0.02$\,\dk,
and this is increased slightly by the inclusion of the four Galactic Centre 
stars. The latter effect is clearly due to the stars showing enhanced 
nitrogen in their atmospheres, but normal oxygen. 
We have previously discussed the possibility of N-rich core gas
contaminating the surface of young massive stars, and hence polluting
the natal photospheric material (RSDR). However due to the lack of 
a strong anti-correlation of C and N, RSDR found no evidence for such
contamination and the same argument holds for the four stars analysed here. 

\begin{figure}
\psfig{file=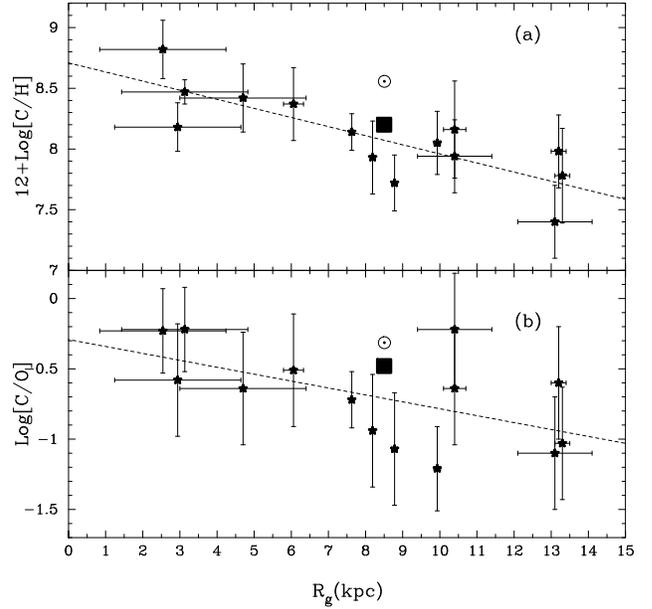,width=9cm}
\caption[]{
{\bf (a):} The Galactic abundance gradient of carbon, composed of the
four programme stars and the data from RSDR. A  gradient of 
$-0.07 \pm 0.02$\,dex\,kpc$^{-1}$ is found, similar to that 
reported by RSDR.
Three of the four stars in our sample do appear C-rich from 
their differential abundances. The solar symbol and the solar
neighbourhood B-type star value (the large square) are taken from 
Table\,\ref{LS5130_abundances}.
{\bf (b):} The $\log$\,C/O abundances are plotted and a least-squares gradient
of $-0.05 \pm 0.02$\,dex\,kpc$^{-1}$ is deduced. This provides stronger
evidence than in RSDR that a C/O gradient exists in the Galaxy and is
due to the four Galactic Centre stars
having normal O abundances, while three are significantly C-rich. The solar 
and neighbourhood B-type star ratios are again taken from 
Table\,\ref{LS5130_abundances}.}
\label{fig_CO_grads}
\end{figure}

An explanation of the N/O gradient was discussed in RSDR, based on the
original ideas of Vila-Costas \& Edmunds (\cite{vil93}). At metallicities 
higher than $12 + \log$ O/H $\sim$8.3, $\log$ N/O would be proportional 
to $\log$ O/H, because N enrichment would be dominated by secondary 
production, i.e. from C and O seed nuclei in the 
natal interstellar material through the CNO-cycle. 
Primary production would play a minimal role in contributing to the ISM 
abundance of N, and simple closed-box models in which the production of 
N is dominated by secondary production do predict a linear
trend. 
At low metallicities the secondary component 
becomes less important (due to its inherent dependence on metallicity), 
and the dominant mechanism for the production of nitrogen 
is primary. The metallicity at which this happens is around
8.3\,dex, close to the value found in the outer most regions of
the Galactic disk. These ideas stem from the fact that 
the N/O ratios in low metallicity dwarf galaxies show
a very large scatter, with little discernable trend. As
primary nitrogen orginates in low-intermediate mass stars 
there is a time delay between is enrichment and that of 
primary oxygen (from short-lived massive stars), hence the 
scatter at low metallicitities is often interpreted as 
time dependent delays between oxygen and nitrogen 
enrichments. However at the high metallicities we are dealing
with, it is likely that secondary production (in massive
stars is dominant), and there would thus be no delay between
N and O production. 
Our N/O ratios toward the Galactic Centre tend to support the
conclusions of RSDR that the N/O gradient in the Galaxy is real, and
due to secondary N production dominating across the disk.

In Fig.\ref{fig_CO_grads} carbon abundances for the four Galactic Centre 
stars and from the RSDR compilation are plotted. As three of 
our new points are C-rich, it is not surprising 
that the abundances appears to steadily increase toward the Centre. 
As discussed for nitrogen, because the stars are not 
rich in oxygen the C/O ratio is greater than normal, hence producing
a larger (and statistically more significant) gradient than that
given in RSDR. As discussed by Maeder (1992) this would be consistent
with C being produced at a greater rate than O in metal-rich stars
(Section\,\ref{discussion_ii}), due to the increased mass-loss in the 
post He-burning phase.

\section{Conclusions}
The analysis of high-quality data for four blue stars lying 
towards the Galactic Centre, and a comparison with young stars
throughout the Galactic disk, leads to the following conclusions
\begin{enumerate}
\item The four stars appear to be normal, young, massive, 
early B-type stars. They lie slightly out of the Galactic plane and
their calculated distances indicate that they are at 
distances from the disk of $0.4 < z < 1.2$\,kpc. A kinematic 
analysis indicates that they were probably born in the disk within 
$2.5-5$\,kpc of the Galactic Centre and then 
subsequently ejected. Comparison with Galactic structure models 
suggests they may have originated in the stellar or 
molecular ring at $R_{\rm g}\simeq3$\,kpc. 
\item A differential line-by-line abundance analysis of the 
four stars with two bright stars near the Sun indicates
a surprising abundance pattern: 
	\begin{itemize}
	\item Two of the stars (LS5130 and LS4419) have oxygen abundances very
	similar to stars in the solar neighbourhood. The other two (LS4784 and
	LS5130) show only marginal evidence for enhanced oxygen. In all cases
	any enhancement is significantly below the $0.3-0.4$\,dex that 
	would be expected given the Galactic position of the stars and 
	the gradient from RSDR. Oxygen is a very
	reliably determined abundance in these particular types of stars. 
	\item The relatively well measured abundances of Si, Al, Mg and S 
	however do indicate that these stars have enhanced abundances in 
 	most other elements. However, the physical cause of the natal
	gas having a ``normal'' oxygen content and enriched $\alpha$-processed
	elements is unclear. The Si and Mg abundances are consistent with the
	continuation of the linear gradient derived by RSDR 
	for stars in the solar neighbourhood and the outer Galaxy.
	\item The four stars also appear rich in both C and N, again 
	in agreement with a continuation of the RSDR gradients towards the 
	Centre. 
	\item We discuss the possible errors in our analysis, and conclude 
	that it is unlikely that these stars have higher oxygen abundances
	than those typically found in the solar neighbourhood. It is also very 
	likely they are rich in all the other elements studied (C, N, Mg, Si, 
        S, Al) although the physical pocesses that produces such a
	pattern is unclear.
	\end{itemize}

\end{enumerate}

\begin{flushleft}
{\bf Appendix A: Table\,9 and 10 are the lists of equivalent widths of all metal lines
measured in the spectra of the Galactic centre stars and the standards 
$\iota$\,Her, and $\gamma$\,Peg. and is available electronically}
\end{flushleft}

\acknowledgements{Spectroscopic data were obtained at the 
Anglo-Australian Observatory in Siding Spring, New South Wales 
and the ESO 3.6m Telescope on La Silla. We also obtained 
photoelectric photometry on the 0.5m in South Africa, and
CCD photometry on the 1.5m Danish Telescope in La Silla. 
We are grateful to the 
staff at all observatories for their assistance.
We are particularly grateful to Dave Kilkenny at SAAO who
provided urgent initial photometry, which 
allowed some of this work to be incorporated into S. Smartt's PhD
thesis. 
Data reduction was performed on the PPARC funded 
Northern Ireland {\small STARLINK} node, and some of the model atmosphere 
programs were made available through the PPARC supported Collaborative 
Computational Project No. 7. We made use of the SIMBAD database 
maintained at CDS, Strasbourg.
We acknowledge financial support for this work from the British 
Council and the Deutscher Akademischer Austauschdienst in the context of the
British-German Academic Research Collaboration initiative, and also 
the Visiting Fellowship program at Queen's University which allowed 
travel for SJS to further discussions. SJS and WRJR acknowledge
funding from the PPARC. KAV 
would like to acknowledge travel and observing support
by ESO, and to thank the Henry Luce Foundation for
research funds through a Clare Boothe Luce professorship
award. }

\newpage

\begin{table}
\caption[]{Appendix: Equivalent widths (m\AA) of the non-diffuse helium lines and the 
metal lines in the stars LS5130, LS4419, LS4784 and the spectroscopic 
standard $\gamma$Peg. Each equivalent width has been assigned error estimates as follows --
a: error less than 10\% -- b: error less than 20\% -- c: error greater than 
20\%. }
\begin{scriptsize}
\begin{tabular}{rcccc} \hline\hline
Species and Line       & LS5130    & LS4419 & LS4784 & $\gamma$Peg  \\
\hline
O\,{\sc ii}   3911.96  &     57a   &   96c  & 76b  & 50a \\
\\	    	          	      	
C\,{\sc ii}   3918.98  &     138a  &  130a  & 261a$^{7}$  & -- \\
N\,{\sc ii}   3919.01  &     	   &        &             & \\
O\,{\sc ii}   3919.28  &  	   &        &             & \\
 \\  	    	          	      	
C\,{\sc ii}   3920.69  &     127a  &  113a  & --        &-- \\
 \\	    	          	      
O\,{\sc ii}   3945.04  &     30b   &   63b  & 58c  & -- \\
  \\	    	          	      	
O\,{\sc ii}   3954.37  &     51a   &  63b  & 74b  & 48a \\
	 \\    	          	      	
He\,{\sc i}  3964.73   &    177a   &  187a  & 188a  & 155a \\
    \\	    	          	      	    	          	      	
O\,{\sc ii}   3982.72  &     29c   &   72b$^{1}$ & 60a$^{1}$ & 31a \\
 	 \\    	          	      	
S\,{\sc iii}  3983.77   &    28c    &  --        &--        & 14a \\
	 \\    	          	      	
N\,{\sc ii}   3995.00  &     102a  &   104a  & 93a  & 78a \\
	 \\    	          	      	
N\,{\sc ii}   4035.08  &     35b   &   46a  & 42c  & 25a \\
	 \\    	          	      	
N\,{\sc ii}   4041.31  &     44b   &  51a  & 60b  & 37a \\
 \\
O\,{\sc ii}   4069.62  &     95a   &   115a  &    97b  & 94 \\
O\,{\sc ii}   4069.89  &     	   &         &        &  \\
	   \\  	          	      	 
O\,{\sc ii}   4072.16  &     62a   &   75a  &  43c  & 65a \\
	 \\    	          	      
C\,{\sc ii}   4074.48  &     48b   &  147a$^{2}$  &   141b$^{2}$  & 42a\\ 
C\,{\sc ii}   4074.85  &           &        &    &         \\
\\
C\,{\sc ii}   4075.85  &           &        &    &          \\
O\,{\sc ii}   4075.86  &     124a  &  --    & --    & --    \\
C\,{\sc ii}   4076.53  &           &        &       &      \\
N\,{\sc ii}   4076.91  &           &        &       &      \\
	 \\    	          	          
O\,{\sc ii}   4078.84  &     24c   &  19c  & --    & 26a \\
 \\
O\,{\sc ii}   4085.11  &     32c   &  39b  & 34b  & 26a \\
	 \\   	          	          
O\,{\sc  ii}   4087.15  &     22c   &  --   & --   & 12a \\
 	 \\    	          	          
Si\,{\sc iv}  4088.85   &    37c    &  44b  &  61a  & --\\
O\,{\sc ii}    4089.28 &  	   &         &     &  \\
    		       		       	 \\\hline
\end{tabular}
\end{scriptsize}
\end{table}

\newpage

\begin{scriptsize}
\begin{tabular}{rcccc} \hline\hline
Species and Line       & LS5130    & LS4419 & LS4784 & $\gamma$Peg  \\
\hline    	          	          

Si\,{\sc iv}  4116.10   &    20c   &  -- &  -- & -- \\
	 \\    	          	       
O\,{\sc ii}   4119.22  &     62a   &  --  & --  & -- \\
   	 \\    	          	               
O\,{\sc ii}   4120.28  &     291a   & -- & --  & --\\
He\,{\sc i}  4120.81   &            &    &    & \\
He\,{\sc i}  4120.99   &            &    &    & \\
O\,{\sc ii}   4121.46  &            &    &    & \\	 	       
   	  \\   	          	       
Si\,{\sc ii}  4128.07   &    53a   &  38c &  65b  & 33a \\
	  \\   	          	       
Si\,{\sc ii}  4130.89   &    54a   &  86c$^{3}$  & 70c$^{3}$ & 34a \\
\\
Ar\,{\sc ii} 4131.71   &     38b   &  --         & --   & 17a \\
      	  \\   	          	       
O\,{\sc ii}   4132.80  &     33b   &  --         & 54c$^{3}$  & 30a \\
       \\      
Al\,{\sc iii}  4149.50 & --            & 56b  & 64a  & -- \\
Al\,{\sc iii}  4150.14 &               &      &      & -- \\
\\          
S\,{\sc ii}   4153.10  &     60a   &  125b  &   105a  &  -- \\
O\,{\sc ii}   4153.30  &  	   &        &      & \\
	\\     	          		 
S\,{\sc ii}   4162.70  &     47b   &  --    &   --    & -- \\ 
  		\\ 		 
Fe\,{\sc iii}    4164.79 &    45c   &  32b  &  --    & \\ 
	\\ 			 
Fe\,{\sc iii}    4166.86 &    26c   &   -- & --    & 13a \\
	\\ 			 
He\,{\sc i}   4168.97    &  99a   &   99a  &  104a  & -- \\
O\,{\sc ii}    4169.22   &          &      &      &  \\
	  \\  		         	  		         
S\,{\sc ii}    4174.24   &   40b   &  28c  &  --    &  -- \\
	  \\  		         	   		         
N\,{\sc ii}    4179.67   &   14c   &  21c  & --    & -- \\
	  \\ 
O\,{\sc ii}    4185.45   &   36b   &  43b  & --    & 30a \\
   	  \\  		         
O\,{\sc ii}    4189.79   &   46b   &   54b &  65b  & -- \\
	 \\ 	       			       
P\,{\sc iii}  4222.15   &    21c   &  20c  & --    & 21a \\
	   \\	          	       
N\,{\sc ii}   4227.74  &     30b   &  55b  & 61b  & 18a \\
	    \\ 	          	       
N\,{\sc ii}   4236.86  &     51a   &  55a  & 66c  & 28a \\
N\,{\sc ii}   4236.98  &          &        &       &\\
\\          	       
N\,{\sc ii}   4241.78  &     49a   &   86a  & 105b  & -- \\
	    \\   	          	       \hline
\end{tabular}
\end{scriptsize}

\begin{scriptsize}
\begin{tabular}{rcccc} \hline\hline
Species and Line       & LS5130    & LS4419 & LS4784 & $\gamma$Peg  \\
\hline 	 	 			         
S\,{\sc iii}  4253.59   &    --    &    118a  &   --    & -- \\
O\,{\sc ii}   4253.90  &          &           & & \\
O\,{\sc ii}   4254.13  &          &           &   &  \\   
 	     \\  	          	       
C\,{\sc ii}   4267.02  &    262a   &  245a & 305a  &   -- \\  
C\,{\sc ii}   4267.27  &           &      &        &   \\
O\,{\sc ii}   4267.71      &       &     &      & \\
    	      \\ 	
O\,{\sc ii}    4275.56   & --    & 97b  &  123a$^{8}$  & -- \\
          	  \\        
O\,{\sc ii}    4277.40   & --    & 72b  & --    & 16a \\
  \\ 
S\,{\sc iii}   4284.99   & --    & --   &  47c  & 28a \\
  \\ 
S\,{\sc ii}   4294.43  &    35b   &  50b  &    37c  & -- \\
O\,{\sc ii}   4294.79  &  	  &       &         &	\\     
   	     \\  	          	       	    	          	       
O\,{\sc ii}   4303.84  &    30c   &   62a  &  54c  & 26a \\
	     \\  	          	       
C\,{\sc ii}   4313.30  &    17c   &   -- &  -- & -- \\
	   \\    	          	       
O\,{\sc ii}   4317.14  &    58a   &   76a  &   64b  & 56a \\
C\,{\sc ii}   4317.26  &         &         &        &  \\
	  \\     	          	       
O\,{\sc ii}   4319.63  &    48a   &    78a  & 63b  & 52a \\
	  \\     	          
O\,{\sc ii}   4325.76  &    40b   &    23b  &  29c  & 20a \\
  	  \\      
O\,{\sc ii}   4345.56  &    30c   &  -- &  -- & 24a \\
  	  \\     	          	      
O\,{\sc ii}   4347.42  &    26c   &  -- &  -- & 21a \\
	   \\    	          	     
Ar\,{\sc ii}  4348.11   &   32c   &  -- &  -- & -- \\
	      \\ 	          	     
O\,{\sc ii}   4349.43  &    66b   &  -- &  -- & 63a \\
	      \\
O\,{\sc ii}   4351.26  &    31c   &  -- &  -- & 41a \\
	  \\     	          	     
O\,{\sc ii}   4366.89  &    60a   & 61a  & 32c  & 48a \\
   	  \\     	          	     
O\,{\sc ii}   4369.27  &    -- &    -- & -- & 14a \\
	  \\     	          	     
C\,{\sc ii}   4372.49  &    52a   & 91a  &   67c$^{9}$ & -- \\
	  \\     	          	     
C\,{\sc ii}   4374.27  &    32b   & -- & -- & -- \\
		  \\ 	     			\hline
\end{tabular}
\end{scriptsize}

\begin{scriptsize}
\begin{tabular}{rcccc} \hline\hline
Species and Line       & LS5130    & LS4419 & LS4784 & $\gamma$Peg  \\
\hline  	          	       
Fe\,{\sc iii}    4395.78  &  40b   &  60a  &  58b  & -- \\
O\,{\sc ii}    4395.94    &        & &             & \\
\\    
C\,{\sc ii}     4411.20  &  38b   & 63b  & -- & -- \\
C\,{\sc ii}     4411.52  &       &       &    & \\
	   \\    		      
O\,{\sc ii}     4414.90  &  84a   & 95a  &    52c  & 75a \\ 
	    \\   		      
O\,{\sc ii}     4416.97  &  64a   & 76a  & 28c  & 61a \\
		  \\ 	     
Fe\,{\sc iii}    4419.59 &    44b & 32c  &  35c  & 34a \\
		  \\ 	     
Fe\,{\sc iii}    4430.95 &  33c   & -- &  39c  &  24a \\
		  \\ 	     
N\,{\sc ii}    4432.74  &  38c   &  -- &  51c  &   20a \\
		  \\ 	     
He\,{\sc i}   4437.55 &  126a   &   103a  & 135a  & 118a \\
		  \\ 	     
N\,{\sc ii}    4442.02 &   -- &    -- &  38c  & 9a \\
		  \\ 	     
N\,{\sc ii}    4447.03  &  60a   & -- &  72b  & 15a \\
O\,{\sc ii}    4448.19  &        &   &       &  \\
		  \\ 	     
O\,{\sc ii}    4452.37  &  31c   &  -- &  29c  & 25a \\
  \\ 
Al\,{\sc iii}   4479.89  &  55a   &  -- &   -- &  38a \\
Al\,{\sc iii}   4479.89  &     &        &       & \\ 
  \\ 
Mg\,{\sc ii}    4481.13 &  202a   & 196a  &  265a  & 160a \\
Mg\,{\sc ii}    4481.33 &      &        &       & \\  
		  \\ 	      
Al\,{\sc iii}   4512.54  &  57a   &  65a  &   63a  & 39a \\
 		  \\ 	      
Al\,{\sc iii} 4528.91 & 73a & 144a$^{10}$ & 135a$^{10}$  & 66a \\
Al\,{\sc iii}   4529.20  &       & &                     & \\
		  \\ 	      
N\,{\sc ii}    4530.40 &   29c   &  -- &      -- & 25a \\
			  \\
N\,{\sc ii}    4552.53 & 174a   &  199a  &   187a  & 114a \\
Si\,{\sc iii}   4552.62 &        & &                &\\
		  \\ 	      
Si\,{\sc iii}   4567.82 &  150a   &  152a  &    151a  & 102a \\ 
  		  \\ 	      
Si\,{\sc iii}   4574.76  &  81a   &  97a  & 90a  &  69a \\
		  \\ 	      
O\,{\sc ii}    4590.97  &  52a   &   87a  & 67a  & -- \\
		  \\ 	      
O\,{\sc ii}    4596.18  &  59a   & 74a  &   52a  & 47a \\
		  \\ 	      
N\,{\sc ii}    4601.48  & 62b   &  107a  &   90a  & 37a \\
O\,{\sc ii}    4602.13  &       &         &        & 15a \\       
  		  \\ 	      \hline
\end{tabular}
\end{scriptsize}

\begin{scriptsize}
\begin{tabular}{rcccc} \hline\hline
Species and Line       & LS5130    & LS4419 & LS4784 & $\gamma$Peg  \\
\hline       
N\,{\sc ii}    4607.16  &  56a   &   64a  & 68a  & 35a \\
		  \\ 	      
O\,{\sc ii}    4609.44  &  41b   &  63a  &  50a  & 30a\\
\\ 	 
N\,{\sc ii}    4613.87  &  55a   &  79a  &  56a  & 27a \\
		  \\ 	      
C\,{\sc ii}    4618.40  &  137c   & 82a  &   213a$^{11}$  & -- \\
C\,{\sc ii}    4619.23 	&         &      &         & \\
    		  \\ 	      
N\,{\sc ii}    4621.29   &  41b   & 73a  & -- & 26a \\
  		  \\ 	      
N\,{\sc ii}    4630.54   &  92a   & 122a  & 123a  & \\
		  \\ 	      
O\,{\sc ii}    4638.86   &  61b   &  81a  &  71a  & 54a \\
		  \\ 	      
O\,{\sc ii}    4641.82   &  74b   & 171a$^{4}$ &   172a$^{4}$   & 76a \\
		  \\ 	      
N\,{\sc ii}    4643.09   &  58b   & -- & -- & 42a \\
		  \\ 	      
C\,{\sc iii}  4647.42  & -- &    230a $^{5}$ &    192a$^{5}$  & --\\
		  \\ 	      
O\,{\sc ii}    4649.14   &  90a   &  -- & -- & 92a \\
		  \\ 	      
O\,{\sc ii}    4650.84  &   64b   &  -- & -- & 52a \\
  \\ 
O\,{\sc ii}    4661.63  &   67a   & 94a  &   75a  & 54a \\
		  \\ 	      
O\,{\sc ii}    4673.74  &   26c   &  -- & -- & 19a \\
		  \\ 	      
O\,{\sc ii}    4676.24  &   44c   &   45c  & -- & 46a \\
			  \\       
N\,{\sc ii}    4678.10  &   24c   & -- & -- & -- \\
			  \\       
N\,{\sc ii}    4694.70  &   15c   & -- & -- & 9a \\
		  \\
O\,{\sc ii}    4696.35  &   15c   & -- & -- & 12a \\
		  \\ 	      
O\,{\sc ii}    4699.00  & 42a   & 69c  &  37a  &  -- \\
O\,{\sc ii}    4699.22  &       &      &       &
		  \\ 	      
O\,{\sc ii}    4703.16  &  --     &  66a$^{5}$  &   53a$^{5}$  & -- \\
  	  \\ 		      
O\,{\sc ii}    4705.35  &   47b   &  --   & --   & -- \\
	  \\ 		      
O\,{\sc ii}    4710.01  &   30b   & 40c  & --   & 20a \\
	  \\ 		      
He\,{\sc i}   4713.14   &  289a   &  283a  & 250a  & 271a \\
He\,{\sc i}   4713.37    &        & &              & \\
	  \\ 		      \hline
\end{tabular}
\end{scriptsize}

\begin{scriptsize}
\begin{tabular}{rcccc} \hline\hline
Species and Line       & LS5130    & LS4419 & LS4784 & $\gamma$Peg  \\
\hline  	      
Si\,{\sc iii}   4716.65   &   41b   & 52b  & --   & 12a \\
	  \\ 		      
N\,{\sc ii}    4779.67  &   18c   & 36c  &  44b  & 10a \\
	  \\ 		      
C\,{\sc ii}    4802.70  &    28c   & --   & --   & 15a \\
	  \\ 		      
N\,{\sc ii}    4803.29  &    37c   & 61a  &   74a  & 18a \\
\\	        
Ar\,{\sc ii}   4806.07  &    29c   & 21c  &  --   & 18a \\
		  \\ 	      
Si\,{\sc iii}   4813.30  &    31b   & 41a  &   75b$^{12}$ & 21a \\
		  \\ 	      
S\,{\sc ii}    4815.52  &    54b   & 27b  &  --   & 27a \\
		  \\ 	      
Si\,{\sc iii}   4819.72  &    55b   & 55a  &    50a  & 40a \\
	  \\ 		      
S\,{\sc ii}    4824.07  &   12c   & -- &  18c  & -- \\
	  \\ 		      
Si\,{\sc iii}   4828.96  &   44b   &  64a  & 59a  & -- \\
	  \\ 		      
Ar\,{\sc ii}   4879.14   &  28b   & -- & 19c  &  -- \\
	  \\ 		      			   
O\,{\sc ii}    4890.93   &  21c   & 22b  & --   & 11a \\
	  \\ 		      
O\,{\sc ii}    4906.83   &  38b   & -- &  29b & -- \\
  \\ 
O\,{\sc ii}    4941.07    & 25b   & 38b  &  14c  &  15a \\
	  \\ 		       
O\,{\sc ii}    4943.00  &   39b   & 52b  &    45c  & 28a \\
	  \\ 		       			       
S\,{\sc ii}    4991.94  &  16c   & -- & 24c  & 22a \\
		  \\	       
N\,{\sc ii}    4994.36   &  31b   & 55a  & 59b  & -- \\
	  \\
S\,{\sc ii}    5009.62   &  32b   & -- & -- & -- \\
		  \\	       
N\,{\sc ii}    5010.62   &  40b   &  40c  & -- & -- \\
	  \\		       
S\,{\sc  ii}    5014.03   &  52b   &  -- & -- & -- \\
	  \\		       
He\,{\sc i}   5015.68   & 322a   &  319a  &  354a  & -- \\
	  \\		       
S\,{\sc ii}    5032.41   &  66b   & 56a  &   91a  & -- \\
		  \\	       			       
Si\,{\sc ii}   5041.03   & 24c   & -- &   30c  & 15a \\
		  \\	   	\hline
\end{tabular}
\end{scriptsize}

\begin{scriptsize}
\begin{tabular}{rcccc} \hline\hline
Species and Line       & LS5130    & LS4419 & LS4784 & $\gamma$Peg  \\
\hline 	           
N\,{\sc ii}    5045.09   &  54b   & 59a  & 47c  &   -- \\
		  \\		    
He\,{\sc i}   5047.74   & 207a   &  163a  & 179b  &   -- \\
		  \\		       
Si\,{\sc ii}   5055.98   &  35c   & -- & 56b  &   22a \\
Si\,{\sc ii}   5056.31   &      &      &      &   \\
\\      
S\,{\sc ii}    5103.30   &   -- &  -- &  20c  & -- \\
		  \\		       
C\,{\sc ii}    5132.96   &  97a   &  74a  &  99a  & -- \\
C\,{\sc ii}    5133.29   &          &      &    &  \\
		  \\		          
O\,{\sc ii}    5159.94   &  23a   &     38a  & 37a  & -- \\
\hline
\end{tabular}

\begin{flushleft}
{\small Notes:\\
1. Blend of O\,{\sc ii} 3982.72 and S\,{\sc iii} 3983.77. \\
2. Includes the O\,{\sc ii}, N\,{\sc ii} and C\,{\sc ii} lines at 
4075--4076\,AA.\\
3. Blended with Ar\,{\sc ii} 4131.71.\\
4. Blended with N\,{\sc ii} 4643.09.\\
5. Blended with O\,{\sc ii} 4649.14 \& 4650.84.\\
6. Blended with O\,{\sc ii} 4705.35.\\
7. Also includes C\,{\sc ii} 3920.69.\\
8. Blended with O\,{\sc ii} 4277.40.\\
9.  Blended with C\,{\sc ii} 4374.27.\\
10. Blended with N\,{\sc ii} 4530.40.\\
11. Blended with N\,{\sc ii} 4621.29 .\\
12. Blended with S\,{\sc ii} 4815.52.\\}
\end{flushleft}
\end{scriptsize}

\newpage
\begin{table}[h]
\caption[]{Equivalent widths (m\AA) of the non-diffuse helium lines and the 
metal lines in the stars LS5127 and $\iota$Her.}
\begin{scriptsize}
\begin{tabular}{rcc} \hline\hline
Species and Line      &  LS5127 & $\iota$Her \\\hline
O\,{\sc ii}  3911.96  &   50c  & 12a \\
\\				
C\,{\sc ii}  3918.98  &  340b  &  72a \\
N\,{\sc ii}  3919.01  &        &  --  \\		
O\,{\sc ii}  3919.28  &        &   8a  \\		     
C\,{\sc ii}  3920.69  &      &  88a \\
\\	
O\,{\sc ii}  3954.37  &   44c  & 20a \\
\\				
He\,{\sc i} 3964.73   & 210c  & 132a \\
\\ 								
N\,{\sc ii}  3995.00  &  108a  & 40a \\
\\								
N\,{\sc ii}  4041.31  &   57b  & 19a \\
\\				
N\,{\sc ii}  4043.53  &   59b  & 9a \\
\\								
O\,{\sc ii}  4069.62  &  182a  & 16a \\
O\,{\sc ii}  4069.89  &        & 19a \\      	
O\,{\sc ii}  4072.16  &        & 27a \\ 		
\\				
C\,{\sc ii}  4074.48  &  185a  &  15a\\
C\,{\sc ii}  4074.85  &        &   9a \\	
O\,{\sc ii}  4075.86  &        &  32a \\	
C\,{\sc ii}  4076.53  &        &  -- \\
N\,{\sc ii}  4076.91  &        & -- \\     	
\\				
O\,{\sc ii}  4119.22  &  324a  & -- \\
O\,{\sc ii}  4120.28  &        &\\	
He\,{\sc i} 4120.81   &        &	\\	
He\,{\sc i} 4120.99   &        &	\\	
O\,{\sc ii}  4121.46   &        &	\\	
Fe\,{\sc iii} 4122.06 &        &  	\\
Fe\,{\sc iii} 4122.98  &        & 	\\
\\
He\,{\sc i} 4168.97  &   70b  &  81a \\
\\				
N\,{\sc ii}  4241.78  &   67a  & 16a \\
\\				
S\,{\sc iii} 4253.59  &   69a  & 13a \\
O\,{\sc ii}  4253.90  &        & 4a \\	
O\,{\sc ii}  4254.13  &        & 2a\\		
\\ 				
C\,{\sc ii}  4267.02  &  338a  & 184a\\
C\,{\sc ii}  4267.27  \\
O\,{\sc ii}  4267.71  &        & \\
Si\,{\sc iii} 4267.80 &        & 26a \\  
\end{tabular}
\end{scriptsize}
\end{table}

\newpage

\begin{scriptsize}
\begin{tabular}{rcc} \hline\hline
Species and Line      &  LS5127 & $\iota$Her \\\hline   	
S\,{\sc ii}  4294.43   &  43a  & 24a \\
O\,{\sc ii}  4294.79   &        & 3a \\ 		
\\			
O\,{\sc ii}  4303.84   &  39a  & 9a \\
\\			
O\,{\sc ii}  4325.76  &   55c  & 6a \\
C\,{\sc ii}  4325.83 	&        & -- \\	
C\,{\sc ii}  4326.16 	&        & --\\	
	\\			
S\,{\sc iii} 4361.53  &   39c  & 4a \\  
\\
O\,{\sc ii}  4366.89  &   42c  & 16a \\
 \\			
C\,{\sc ii}  4372.49  &   48c  & -- \\
\\			
Fe\,{\sc iii} 4395.78  &   43b & 15a \\
\\			
O\,{\sc ii}  4414.90  &  101b &  28a \\
O\,{\sc ii}  4416.97   &        & 23a \\
\\
Fe\,{\sc iii} 4419.59  &   53a  & 23a \\
\\			     
He\,{\sc i} 4437.55  &  101a  & 102a \\
\\			       			       
N\,{\sc ii}  4447.03  &   65b  & 18a \\
\\
S\,{\sc ii}  4463.58  &   36b  & 14a \\
\\
Mg\,{\sc ii} 4481.13  &  423a  & 205a \\
Mg\,{\sc ii} 4481.33  &        & \\ 	       
\\			      		       
Al\,{\sc iii} 4512.54  &   60a & 20a \\ 
\\
S\,{\sc ii}  4524.95   &  47b  & 22a \\
\\
Al\,{\sc iii} 4528.91  &   100b & 6a \\
Al\,{\sc iii} 4529.20   &        & 28a\\       
N\,{\sc ii}  4530.40    &        & 7a \\       
\\	 \hline
\end{tabular}
\end{scriptsize}

\newpage
\begin{scriptsize}
\begin{tabular}{rcc} \hline\hline
Species and Line      &  LS5127 & $\iota$Her \\\hline       
Fe\,{\sc ii} 4549.47   &   49b  & -- \\
\\ 			       
S\,{\sc ii}  4552.38   &  191a  & 71a \\
Si\,{\sc iii} 4552.62   &        & \\       
\\		       		       
Si\,{\sc iii} 4567.82  &   156a & 59a \\
\\
Si\,{\sc iii} 4574.76   &   93a & 27a \\
\\
Fe\,{\sc ii} 4583.83   &   39b  & 12a \\
\\
O\,{\sc ii}  4590.97   &   39c  & 17a \\
\\
O\,{\sc ii}  4596.18    & 32c  & 12a \\
\\		       
N\,{\sc ii}  4601.48    &  49c  & 15a \\
\\			       
N\,{\sc ii}  4607.16   &   47c  & 14a \\
\\       
C\,{\sc ii}  4618.40   &  114c  & -- \\
C\,{\sc ii}  4619.23   &        & \\      
N\,{\sc ii}  4621.29   &        & \\         
Si\,{\sc ii} 4621.42     &        & \\       
Si\,{\sc ii} 4621.72     &        & \\       
\\
Fe\,{\sc ii} 4629.34  &   107a  &   -- \\
N\,{\sc ii}  4630.54   &        & \\        	 
\\
O\,{\sc ii}  4638.86  &    43a  & 21a \\
 \\
O\,{\sc ii}  4641.82   &  108a  & 24a \\
N\,{\sc ii}  4643.09   &        &  17a \\
\\
S\,{\sc ii}  4648.17   &  170a  &  5a \\
O\,{\sc ii}  4649.14   &        &  34a \\     	 
O\,{\sc ii}  4650.84   &        &  17a \\ 
\\ 
S\,{\sc ii}  4656.74   &   26b  & 10a \\
\\
O\,{\sc ii}  4661.63   &   65b  &  15a \\  
Al\,{\sc iii} 4663.05  &        & 13a \\
\\
He\,{\sc i} 4713.14   &  306a  &  239a \\
He\,{\sc i} 4713.37   &        & \\
\\      
Si\,{\sc iii} 4813.30  &    81b  & 8a \\
S\,{\sc ii}  4815.52   &        &  40a \\      	 
\\
Si\,{\sc iii} 4819.72   &   47c  & -- \\
\\				 \hline
\end{tabular}
\end{scriptsize}

\newpage
\begin{scriptsize}
\begin{tabular}{rcc} \hline\hline
Species and Line      &  LS5127 & $\iota$Her \\\hline 

S\,{\sc ii}  5014.03   &  522a  &  38a \\
He\,{\sc i} 5015.68    &        & 233a \\     	 	 
\\
S\,{\sc ii}  5032.41  &   128a  & 47a \\
\\
Si\,{\sc ii} 5041.03   &   85a  &  42a \\
\\ 
N\,{\sc ii}  5045.09  &   190a  & \\
S\,{\sc ii}  5047.28   &        & \\      	 
He\,{\sc i} 5047.74    &        & \\     	 
\\			 
Si\,{\sc ii} 5055.98   &  182a  & 82a  \\
Si\,{\sc ii} 5056.31   &        & \\
\\\hline
\end{tabular}
\end{scriptsize}

\end{document}